\documentclass[12pt]{article}

\usepackage{setspace} 
\usepackage[ansinew]{inputenc} % Input
\usepackage{amsthm}
\usepackage{mathtools}

\usepackage[T1]{fontenc} % Font encoding
\usepackage{lmodern,microtype} % Typeface

\usepackage{titlesec,titling} % Section titles
\usepackage[nohead]{geometry} % Page & margins
\usepackage{setspace} % Spacing
\usepackage{enumitem,booktabs} % Tables & lists
\usepackage{amsmath,amsfonts,amssymb,amsthm}
\usepackage{mathrsfs,ushort} % Math style
\usepackage{graphicx,psfrag,epsf}
\usepackage{enumerate}
\usepackage{natbib}
\usepackage{bm}
\usepackage{dsfont}
\usepackage{url} % not crucial - just used below for the URL 

\usepackage{hyperref}
\usepackage{bm}
\usepackage{mathtools}
\usepackage{dsfont}
\usepackage{comment}
\usepackage{breqn}
\usepackage{colonequals}
\usepackage{mathtools}
\usepackage{cases}
\usepackage{threeparttable}
\usepackage{mathtools}
\usepackage{graphicx}
\usepackage{empheq}
\usepackage{lmodern}
\usepackage{amsmath}

\usepackage{colonequals}
\RequirePackage{natbib}

\newcommand\Item[1][]{%
	\ifx\relax#1\relax  \item \else \item[#1] \fi
	\abovedisplayskip=0pt\abovedisplayshortskip=0pt~\vspace*{-\baselineskip}}

\titleformat{\section}[block]{\centering\large\bfseries}{\thesection.}{0.5em}{}
\titleformat{\subsection}[block]{\flushleft\bfseries}{\thesubsection.}{0.5em}{}
\titleformat{\subsubsection}[runin]{\normalsize\itshape}{\bfseries\thesubsubsection.}{0.5em}{}[.--\:]

\titlespacing{\section}{0ex}{6ex}{3ex}
\begin{document}

\def\spacingset#1{\renewcommand{\baselinestretch}%
{#1}\small\normalsize} 
%\spacingset{1.28}

%%%%%%%%%%%%%%%%%%%%%%%%%%%%%%%%%%%%%%%%%%%%%%%%%%%%%%%%%%%%%%%%%%%%%%%%%%%%%%

%\if0\blind
%{
  \title{\bf Direct Multi-Step Forecast based Comparison of Nested Models via an Encompassing Test}
  \author{ Jean-Yves Pitarakis\thanks{I wish to thank the ESRC for its financial support via grant ES/W000989/1.  
  		Address for Correspondence: Jean-Yves Pitarakis, Department of Economics, University of Southampton, Southampton SO17 1BJ, United-Kingdom. 
  		%Matlab codes and replication files are available from %\url{https://github.com/jpitarakis/Multi-Step_Encompassing.git}
  		  	} \\
    Department of Economics \\
    University of Southampton}
  \maketitle
%} \fi

%\if1\blind
%{
%  \bigskip
%  \bigskip
%  \bigskip
%  \begin{center}
%    {\LARGE\bf Title}
%\end{center}
%  \medskip
%} \fi

\bigskip
\begin{abstract}
We introduce a novel approach for comparing out-of-sample multi-step forecasts obtained from a pair of nested models that is based on the forecast encompassing principle. Our proposed approach relies on an alternative way of testing the population moment restriction implied by the 
forecast encompassing principle and that links the forecast errors from the two competing models in a particular way. Its key advantage is that it is able to bypass the variance degeneracy problem afflicting model based forecast comparisons across nested models. It results in a test statistic whose limiting distribution is standard normal and which is particularly simple to construct and can accommodate both single period and longer-horizon prediction comparisons. Inferences are also 
shown to be robust to different predictor types, including stationary, highly-persistent and purely deterministic 
processes. Finally, we illustrate the use of our proposed approach through an empirical application that explores the role of global inflation in enhancing individual country specific inflation forecasts.  
\end{abstract}

\noindent%
\vfill

\newpage

\spacingset{1.4888} % DON'T change the spacing!

\section{Introduction}

Comparing the out-of-sample predictive ability of alternative models is an essential component of empirical research. Whether one wishes to evaluate the quality of a model based forecast relative to one obtained from a simpler benchmark for the purpose of validating a particular theory or to simply select the model that produces the most accurate forecast of some outcome of interest, one needs to look beyond in-sample specification tests by also assessing the comparative out-of-sample accuracy of the forecasts produced by competing and often nested models. Our goal in this paper is to propose a novel approach to testing whether multi-step forecasts from one model encompass the forecasts from a rival larger model when the two models have a nested structure. Such an encompassing based approach offers an intuitive way of comparing competing forecasts as it is designed to assess whether or not a larger model contains relevant predictive information about an outcome of interest not already captured by a smaller specification. 

In this paper we propose an alternative way of testing a particular population moment restriction implied by the forecast encompassing principle and which links the forecast errors obtained from two candidate models in a particular way. The novelty lies in its ability to bypass the variance degeneracy problem afflicting
nested model based forecast comparisons without incurring any data loss. Such degeneracy naturally arises in nested model comparison settings as under the null hypothesis of equal predictive accuracy in the population the two models are identical and this results in quantities such as MSE spreads and their variance 
collapsing to zero asymptotically. 

The objective of comparing out-of-sample forecasts has been at the centre of a vast literature inspired by the early work in \cite{dm1995} and \cite{w1996} who developed a theory for testing 
whether two alternative forecasts have an equal population level predictive ability under some given loss function (e.g., via mean squared prediction errors). The literature that followed focused on variations of these tests designed to broaden their scope and applicability, relax the assumptions under which their asymptotic properties hold etc. (see \cite{w2006} and \cite{cm2013} for a comprehensive overview of this agenda). In parallel to this literature an alternative approach aimed at comparing the relative forecasting ability of two competing models relied on the so-called forecast encompassing principle developed in the early work of \cite{hr1982}, \cite{mr1986}, \cite{ch1986} amongst others. In its simplest formulation this principle 
is based on the observation that when one of two forecasts does not contain any useful information not already present in the rival forecast, combining the two forecasts via an optimal convex combination 
will not result in a smaller squared error loss. Within a nested and model based forecast setting this translates into the optimal combined forecast assigning zero weight to one of the forecasts. Accordingly, we say that the forecasts based on the larger model are encompassed in the forecasts of the smaller model. 
The forecast encompassing literature has considered numerous and often closely related regression based methods for implementing such tests. These often reduce to testing the validity of a population moment restriction
under which the optimal combination weight associated with the encompassed forecast is zero (see \cite{hln1998}, \cite{ch2009} for an overview of the early developments in this literature and \cite{bk2021}, \cite{ors2023} for more recent implementations and uses of the encompassing principle). 

An important challenge for either of these two approaches to out-of-sample model based forecast evaluation and comparison stems from a variance degeneracy problem affecting inferences in nested environments where one typically 
has a parsimonious benchmark model that is being compared with a competitor model containing additional predictors. Under the null hypothesis of equal predictive accuracy for instance (e.g., equality of population MSEs) the two nested models will be equivalent in the sense of sharing the same 
forecast errors asymptotically and in the limit both the sample MSE differentials and their variance 
will thus equal zero, invalidating the use of DM type studentized statistics for inference purposes. A similar complication also arises when testing the null hypothesis that the larger model's forecasts are encompassed in the smaller model's forecasts when the two models are nested. Under this null hypothesis it is again the  case that the additional predictors of the larger model do not convey any useful MSE-reducing information not already present in the smaller model. %Under either the null hypothesis of equal predictive accuracy or the null hypothesis of encompassing the two models being compared are equivalent and as a result the variance of their squared out-of-sample forecast error differences is identically equal to zero invalidating inferences based on studentized statistics. \\

These difficulties have led to an important and widely referenced agenda that sought to address the variance degeneracy problem by considering alternative normalisations 
of MSE loss differentials so as to obtain well defined null asymptotics for DM as well as encompassing based test statistics (see \cite{cm2001,cm2005, cm2013}), \cite{m2007}, \cite{ht2015} amongst others). 
A practical limitation of these methodological developments is the highly non-standard nature of the  asymptotics of these test statistics, typically expressed in terms of functionals of stochastic integrals in Brownian Motions. Moreover, these limiting distributions also depend on a large number of model specific nuisance parameters which render their practical implementation difficult, often requiring simulation based approaches. This is the case for instance when one considers forecast horizons greater than one and/or when the models are characterised by conditional heteroskedasticity. More recently, with a focus on one-step ahead forecasts \cite{pit2023} proposed a novel approach to the implementation of DM type tests in nested settings in a way that bypasses the variance degeneracy problem while at the same time maintaining convenient standard normal asymptotics. 
The theory developed in \cite{pit2023} relied on the observation that the averaged squared forecast errors of two models 
need not be compared over the full effective sample span. It can instead be compared over non-identical but overlapping sample segments. This simple modification was shown to ensure well defined standard normal asymptotics for DM type constructions at the expense of very little data loss. 

Our goal in this paper is to consider a multi-step nested model based forecasting environment and propose a new approach for conducting forecast encompassing tests (as opposed to equal predictive ability tests) in a way that avoids the variance degeneracy problem plaguing nested model comparisons. The way we propose to bypass the variance degeneracy problem in this encompassing testing framework is fundamentally different from the methods developed in \cite{pit2023}) as our approach does not require any data discarding
and is based on an alternative approach that requires testing the validity of a population moment restriction. A further novelty of this paper is its ability to handle long-horizon based forecast comparisons which are dealt with via local projection type regressions.  

Here it is also important to explicitly point out that our analysis is concerned with population level predictive accuracy comparisons with forecasts being generated recursively via an expanding window approach. 
It is precisely in such an environment that nested model structures cannot be handled using existing tools. This contrasts with finite sample based predictive accuracy comparisons as for instance in \cite{gw2006}. These latter approaches must rely on rolling-window based forecasting schemes so as to avoid the complications caused by competing models collapsing to the same model asymptotically. 

The plan of the paper is as follows. Section 2 introduces our nested modelling framework, forecasting setup and hypotheses of interest. Section 3 introduces our novel approach to the implementation of forecast encompassing tests. Sections 4 and 5 focus on their null asymptotics and local power properties respectively. Section 6 assesses their finite sample size and power properties via a comprehensive simulation exercise. Section 7 illustrates the usefulness of the proposed techniques via an empirical application which highlights the important role played by global factors for enhancing country specific inflation forecasts. Section 8 concludes with a summary of our key findings and suggestions for further research. 

\section{Forecasting Framework and Hypotheses}

Given observations $(y_{t},{\bm x}_{1,t}^\top,{\bm x}_{2,t}^\top)$ for $t=1,\ldots,T$ we wish to retrospectively 
compare the relative accuracy of {\it direct} h-steps ahead forecasts of $y_{t}$ formed from the two competing models
\begin{align}
	y_{t+h} & = \alpha_{01}+{{\bm \beta}}^\top {\bm x}_{1,t}+w_{t+h} \label{eq:eq1} \\
	y_{t+h} & = \alpha_{02}+{{\bm \delta}}^\top {\bm x}_{1,t}+{{\bm \gamma}}^\top {\bm x}_{2,t}+u_{t+h}, \label{eq:eq2}
\end{align}
\noindent
where the ${\bm x}_{i,t}$'s are $(p_{i}\times 1)$ predictor vectors, ${\bm \beta}$ and ${\bm \delta}$ are the $(p_{1}\times 1)$ slope parameter vectors associated with ${\bm x}_{1,t}$, ${\bm \gamma}$ is a $(p_{2}\times 1)$ slope parameter vector associated with ${\bm x}_{2,t}$, and $w_{t}$ and $u_{t}$ are random disturbance terms. 
We also let $\widetilde{\bm x}_{1,t}=(1,{\bm x}_{1,t}^{\top})^{\top}$ and $\widetilde{\bm x}_{12,t}=(1,\bm x_{1,t}^{\top},{\bm x}_{2,t}^{\top})^{\top}$ so that models (\ref{eq:eq1}) and (\ref{eq:eq2}) can be reformulated as 
$y_{t+h}={\bm \theta}_{1}^{\top} \ \widetilde{\bm x}_{1,t}+w_{t+h}$ 
and
$y_{t+h}={\bm \theta}_{2}^{\top} \ \widetilde{\bm x}_{12,t}+u_{t+h}$ respectively with  ${\bm \theta}_{1}=(\alpha_{01},{\bm \beta}^{\top})^{\top}$ and ${\bm \theta}_{2}=(\alpha_{02}, {\bm \delta}^{\top},{\bm \gamma}^{\top})^{\top}$. We take $h\geq 1$ throughout.

 % and where ${\bm \theta}=\arg \min_{{\bm c} \in \mathds{R}^{p}} \sum_{t}E[y_{t+h}-{\bm c}^{\top} {\widetilde{\bm x}}]^{2}$. 
 
 We are interested in retrospectively comparing these two nested models on the basis of their out-of-sample population level relative predictive accuracy whereby forecast errors are evaluated at
 the underlying population parameters. Our approach is based on the forecast encompassing principle whereby h-steps ahead forecasts from model (\ref{eq:eq1}) are said to encompass forecasts from model (\ref{eq:eq2}) if the latter do not add any useful information to the former. Equivalently, if the optimal combination of the two forecasts assigns zero weight to model (\ref{eq:eq2}). 
 
 Throughout this paper  these h-steps ahead forecasts are formed recursively via an expanding window approach starting from a given sample location denoted $k_{0}=[T\pi_{0}]$ for some fixed and given $\pi_{0} \in (0,1)$. This ensures that there are at least $k_{0}$ observations when starting the recursive estimation.  
 Formally, given the observed data $\{y_{t},{\bm x}_{1t},{\bm x}_{2t}\}_{t=1}^{T}$ pseudo out-of-sample forecasts for time period $(t+h)$ from models (\ref{eq:eq1}) and (\ref{eq:eq2}) are obtained as 
  $\hat{y}_{1,t+h|t}=\hat{\bm \theta}_{1,t}^{\top} \ \widetilde{\bm x}_{1,t}$ and
 $\hat{y}_{2,t+h|t}=\hat{\bm \theta}_{2,t}^{\top} \ \widetilde{\bm x}_{12,t}$ where 
 $\hat{\bm \theta}_{1,t}=\left(
 \sum_{s=1}^{t} \widetilde{\bm x}_{1,s-h} \widetilde{\bm x}_{1,s-h}^{\top}\right)^{-1} \sum_{s=1}^{t} \widetilde{\bm x}_{1,s-h} y_{s}$
 and 
 $\hat{\bm \theta}_{2t}=\left(
 \sum_{s=1}^{t} \widetilde{{\bm x}}_{12,s-h} \tilde{{\bm x}}_{12, s-h}^{\top}\right)^{-1} \sum_{s=1}^{t} \tilde{{\bm x}}_{12, s-h} y_{s}$
 for $t=k_{0},\ldots,T-h$. The h-steps ahead forecast errors associated with these two forecasts 
 are $\hat{e}_{1,t+h|t}=y_{t+h}-\hat{y}_{1,t+h|t}$ and
 $\hat{e}_{2,t+h|t}=y_{t+h}-\hat{y}_{2,t+h|t}$. These $n \equiv (T-h-k_{0}+1)$ recursively estimated pseudo 
 out-of-sample h-steps ahead forecast errors form the basis of the encompassing based tests developed below. For later use we also let $\textrm{MSE}_{i,h}=\sum_{t=k_{0}}^{T-h}\hat{e}_{i,t+h|t}^{2}/n$ (i=1,2) denote the sample MSEs associated with the two forecasts. \\
 
 \noindent
 REMARK 1: Specifications such as (\ref{eq:eq1}) and (\ref{eq:eq2}) are often referred to as {\it direct} multi-step forecasting regressions as they model the future h-steps ahead magnitudes of an outcome of interest as a function of current predictors. In this sense they bear strong resemblance with the regressions considered in the local projection literature (see e.g., \cite{jorda2005}, \cite{mp2021}). More specifically we view (\ref{eq:eq1})-(\ref{eq:eq2}) as fitted models used to generate forecasts as opposed to 
 referring to them as competing DGPs. For this reason both ${\bm \theta}_{1}$ and ${\bm \theta}_{2}$ should be made dependent on $h$, a detail we have omitted for notational parsimony (e.g., the slope coefficients 
 in (\ref{eq:eq1}) are such that ${\bm \beta}_{h}=\arg \min_{z \in \mathds{R}^{p_{1}}}E((y_{t+h}-E[y_{t+h}])-z^{\top}({\bm x}_{1,t}-E[{\bm x}_{1,t}]))^{2}$ and similarly for the remainder parameters). To further illustrate these points, suppose for instance that the DGP is given by 
 the simple AR(1) process $y_{t}=\rho y_{t-1}+\epsilon_{t}$ with serially uncorrelated disturbances and consider the counterpart to (\ref{eq:eq1}) for $h=2$. Straightforward algebra gives $y_{t+2}=\rho^{2} \ y_{t}+w_{t+2}$ with 
 $w_{t+2}=\rho \ \epsilon_{t+1}+\epsilon_{t+2}$ i.e., $w_{t} \sim MA(1)$. Within our notation in (\ref{eq:eq1}) 
 we have $\alpha_{01} = 0$ and ${\bm \beta}_{1} = \rho^{2}$ and $w_{t+2}=\rho \ \epsilon_{t+1}+\epsilon_{t+2}$. Similarly, for $h=3$ the parameters of the fitted model (\ref{eq:eq1}) 
 are such that $\alpha_{01}= 0$, ${\bm \beta}_{1} = \rho^{3}$ and 
 $w_{t+3}=\rho^{2}\epsilon_{t+1}+\rho \epsilon_{t+2}+\epsilon_{t+3}$. \\
  
 The difficulty of implementing equal predictive accuracy or forecast encompassing tests in such nested
 settings stems from the variance degeneracy problem that plagues the null asymptotics 
 of any test statistic built on the difference $\Delta \textrm{MSE}_{h}=\textrm{MSE}_{1,h}-\textrm{MSE}_{2,h}$. 
  Conventional normalisations applied to this MSE spread as for instance in a Diebold-Mariano type formulation such as $\textrm{DM}_{n}=\sqrt{n} \ \Delta \textrm{MSE}_{h}/\hat{\omega}_{n}$ or other related formulations for testing $H_{0}\colon E[\hat{e}_{1,t+h|t}^{2}(\bm \theta_{1}^{0})]=E[\hat{e}_{2,t+h|t}^{2}(\bm \theta_{2}^{0})]$
  result in degenerate asymptotics. Note that our formulation of the null hypothesis is meant to emphasise the fact that we are interested in population level predictive ability so that forecasts are compared at the corresponding population parameters. In what follows we omit this detail for notational simplicity.  
  Depending on the chosen long-run variance normaliser $\hat{\omega}_{n}^{2}$, under the null hypothesis either  the numerator or both the numerator and denominator of $\textrm{DM}_{n}$ converge to zero, invalidating the use of such methods in nested settings. These observations have led to a vast literature 
that introduced alternative normalisations whereby quantities such as $n \ \Delta \textrm{MSE}_{n}/\hat{\omega}_{n}$ as opposed to $\sqrt{n} \ \Delta \textrm{MSE}_{h}/\hat{\omega}_{n}$ were shown to lead to well-defined albeit non-standard asymptotics under the null hypothesis (see e.g., \cite{cm2001,cm2005}, \cite{m2007}, \cite{ht2015} and references therein).   

  Similar difficulties also affect encompassing based approaches. 
    These typically reduce to testing the null hypothesis 
  \begin{align}
  	\textrm{E}[\hat{e}_{1,t+h|t}(\hat{e}_{1,t+h|t}-\hat{e}_{2,t+h|t})] & =0 
  	\label{eq:eq3}
  \end{align}
   via statistics of the type $\textrm{ENC}_{n}=\sqrt{n}(\sum \hat{e}_{1,t+h|t}(\hat{e}_{1,t+h|t}-\hat{e}_{2,t+h|t})/n)/\hat{\omega}_{n}$. 
   To highlight the intuition behind (\ref{eq:eq3}) it is useful to recall that the linearly combined forecast $\hat{y}_{c,t+h|t}=(1-\lambda) \ \hat{y}_{1,t+h|t}+\lambda \ \hat{y}_{2,t+h|t}$ for $0\leq \lambda \leq 1$ can be re-expressed in terms of the corresponding forecast errors as $\hat{e}_{c,t+h|t}=\hat{e}_{1,t+h|t}+\lambda \ (\hat{e}_{2,t+h|t}-\hat{e}_{1,t+h|t})$ or $\hat{e}_{1,t+h|t}=\lambda \ (\hat{e}_{1,t+h|t}-\hat{e}_{2,t+h|t})+\hat{e}_{c,t+h|t}$ so that under $\lambda=0$ which is implied by (\ref{eq:eq3}) the forecast $\hat{y}_{2,t+h|t}$ contains no useful information not already present in $\hat{y}_{1,t+h|t}$. Note also that under the alternative whereby the additional predictors in (\ref{eq:eq2}) do have predictive power the combination weight $\lambda$ equals one so that the covariance in (\ref{eq:eq3}) will be positive and the test is therefore one-sided to the right.  In nested settings, statistics such as $\textrm{ENC}_{n}$ or their variants will continue to suffer from the degeneracy problem since asymptotically the two models will share the same forecast errors leading to outcomes such as 
   $\textrm{ENC}_{n}\stackrel{p}\rightarrow 0$ or possibly  $\textrm{ENC}_{n}\rightarrow 0/0$ depending on the chosen long-run variance normalizer. As it was the case for the $\textrm{DM}_{n}$ statistic, alternative normalisations applied to $\textrm{ENC}_{n}$ type statistics also allow the handling of encompassing tests in nested settings but again with non-standard distributions. Moreover, these depend on numerous nuisance parameters
   which are also difficult to {\it clean out} via HAC type scalings when one wishes to consider multi-step as opposed to one-step ahead forecasts. In \cite{cm2001,cm2005,cm2013} for instance the authors considered a variety of encompassing statistics referred to as \textrm{ENC-T}, \textrm{ENC-REG} or \textrm{ENC-NEW} equivalent to variations of $\textrm{ENC}_{n}$ above across alternative choices of $\hat{\omega}_{n}^{2}$.  
   
  Given the above background our main objective here is to develop a new approach for conducting forecast encompassing tests that results in statistics having standard normal and nuisance parameter free null asymptotics even under the nested setting of (\ref{eq:eq1})-(\ref{eq:eq2}). The way we propose to bypass the variance degeneracy problem here is fundamentally different from the methods developed in \cite{pit2023} and which were concerned with predictive accuracy as opposed to encompassing tests. Indeed, our approach does not require any data discarding or the evaluation of $\textrm{MSE}_{1,h}$ and $\textrm{MSE}_{2,h}$ over different subsets of the sample in addition to being able to accommodate multi-step forecasts. Our novel proposal relies on an alternative implementation of the sample counterpart of the population moment restriction in (\ref{eq:eq3}) under which model (\ref{eq:eq2}) does not lead to any squared error loss based forecast quality improvement over model (\ref{eq:eq1}).  Two additional advantages of our proposed approach are its ability to handle serial-correlation and conditional heteroskedasticity in a straightforward manner and its robustness to the presence of highly persistent predictors and/or deterministic trends in (\ref{eq:eq1})-(\ref{eq:eq2}), thus considerably expanding its relevance for applications.

 %Such models are particularly relevant when it comes to testing competing economic theories which can often be reduced for instance to testing the martingale difference null against predictability induced by an additional set of economically meaningful covariates. Also, and more generally, it is often the case that one wishes to assess whether an additional set of predictors improves the forecasting performance of a model. 

\section{A novel formulation of forecast encompassing tests}

The conventional approach for testing the moment restriction in (\ref{eq:eq3}) is to base inferences on a suitably standardized version of its sample counterpart 
\begin{align}
\widehat{d}_{n} & = \frac{1}{n} \sum_{t=k_{0}}^{T-h} \hat{e}_{1,t+h|t}^{2}-\frac{1}{n}\sum_{t=k_{0}}^{T-h} \hat{e}_{1,t+h|t}\hat{e}_{2,t+h|t}  \label{eq:eq4}
\end{align} 
and evaluate the proximity of the latter to zero. The reason such an approach is problematic in nested 
settings is because under the null hypothesis of interest the two sample averages that make up (\ref{eq:eq4}) have the same variance as $n$ or $T$ tend to infinity (recall that throughout this paper $n \equiv  T-k_{0}-h+1$ refers to the effective number of forecast errors and $h$ is treated as fixed so that we may refer to $n$ or $T$ as diverging to infinity interchangeably). 

Here we propose to operationalise the sample counterpart of the population moment restriction of interest differently from (\ref{eq:eq4}). The idea is based on the observation that the classical sample mean is only one among numerous alternative ways of estimating an underlying population mean. By using an alternative estimator of the moment $E[\hat{e}_{1,t+h|t}\hat{e}_{2,t+h|t}]$ in $(E[\hat{e}_{1,t+h|t}^{2}]-E[\hat{e}_{1,t+h|t}\hat{e}_{2,t+h|t}])$, and designed in a way such that it has 
a different limiting variance than the classical sample mean used to estimate $E[\hat{e}_{1,t+h|t}^{2}]$ (while maintaining its unbiasedness and/or consistency) should allow us to bypass the variance degeneracy problem afflicting inferences. 

To achieve the above goal we consider an alternative to the classical sample mean given by the 
linear combination of two subsample means. Generically, instead of computing say $\overline{X}_{1\colon n}=\sum_{t=1}^{n}X_{t}/n$ as an estimator of $E[X_{t}]$ we consider using $\overline{X}_{s}=(\overline{X}_{1\colon m}+\overline{X}_{m+1 \colon n})/2$ with $\overline{X}_{1\colon m}=\sum_{t=1}^{m}X_{t}/m$ and 
$\overline{X}_{m+1 \colon n}=\sum_{t=m+1}^{n}X_{t}/(n-m)$ for some $m \in [1,n]$. Note of course that $\overline{X}_{s}$ reduces to the classical sample mean $\overline{X}_{n}$ if $m=n/2$. This split-sample based estimator shares similar properties to $\overline{X}_{1\colon n}$
and its variance will also be very close to that of $\overline{X}_{1 \colon n}$ provided that $m$ is not chosen too far from $n/2$. The fact that $\overline{X}_{1\colon n}$ and $\overline{X}_{s}$ have different variances is precisely what we need to bypass the variance degeneracy problem afflicting nested model comparisons. What we propose to do therefore is to replace the second component in the right hand side of (\ref{eq:eq4}) by a sample-split based average instead of the classical sample mean. Here, it is particularly important to emphasise that such an approach does not entail any data loss as both $\overline{X}_{1\colon n}$ and $\overline{X}_{s}$ make use of the full available data. The use of such split-sample averages has been advocated in various different contexts e.g., \cite{dh2014} for confidence interval construction, \cite{gp2023} for covariate screening in high dimensional settings.  

Formally, instead of testing $E[\hat{e}_{1,t+h|t}(\hat{e}_{1,t+h|t}-\hat{e}_{2,t+h|t})]=0$ using (\ref{eq:eq4}), we propose to test the same moment restriction using the following generalization of (\ref{eq:eq4}) 
\begin{align}
\widehat{d}_{n}(m_{0}) & = \frac{\sum_{t=k_{0}}^{T-h} \hat{e}_{1,t+h|t}^{2}}{n}-\frac{1}{2}\left(
		\frac{\sum_{t=k_{0}}^{m_{0}+k_{0}-1}\hat{e}_{1,t+h|t} \hat{e}_{2,t+h|t}}{m_{0}}
		+\frac{\sum_{t=m_{0}+k_{0}}^{T-h}\hat{e}_{1,t+h|t} \hat{e}_{2,t+h|t}}{n-m_{0}}\right)
	\label{eq:eq5}
\end{align} 

\noindent
which makes use of a subsample average based estimator for estimating the population mean of $\hat{e}_{1,t+h|t}\hat{e}_{2,t+h|t}$. Note that with $m_{0}=n/2$ - assuming an even number of observations for the exposition - $\widehat{d}_{n}(m_{0})$ in (\ref{eq:eq5}) reduces to $\widehat{d}_{n}$ in (\ref{eq:eq4}) bringing us back to the conventional way of specifying encompassing principle based statistics and which are unsuitable for nested model comparisons. This $m_{0}=n/2$ scenario will naturally be ruled out in our analysis. Recalling that we have $n \equiv T-h-k_{0}+1$ effective forecast errors, in what follows we let $m_{0}=[n \mu_{0}]$ for $0<\mu_{0}<1$ and refer to $m_{0}$ and $\mu_{0}$ as the sample split location and fraction respectively. The requirement that $m_{0}\neq n/2$ naturally translates into $\mu_{0} \neq 1/2$. 
 
Given the above analysis we propose to implement forecast encompassing based tests using the following 
statistic 
\begin{align}
&\resizebox{0.92\hsize}{!}{$
	{\cal E}_{n}(m_{0}) = \dfrac{1}{\widehat{\omega}_{n}} 
	\left(
	\dfrac{\sum_{t=k_{0}}^{T-h}\hat{e}_{1,t+h|t}^{2}}{\sqrt{n}}-\dfrac{1}{2}\left(
	\dfrac{n}{m_{0}}\dfrac{\sum_{t=k_{0}}^{m_{0}+k_{0}-1} \hat{e}_{1,t+h|t} \hat{e}_{2,t+h|t}}{\sqrt{n}}
	+\dfrac{n}{n-m_{0}}\dfrac{\sum_{t=m_{0}+k_{0}}^{T-h} \hat{e}_{1,t+h|t} \hat{e}_{2,t+h|t}}{\sqrt{n}}\right)
	\right)$}
\label{eq:eq6}
\end{align}
\noindent
where $\widehat{\omega}_{n}^{2}$ generically denotes a suitable variance normalizer. 
It is also convenient to express (\ref{eq:eq6}) more compactly as the studentized sample mean 
\begin{align}
	{\cal E}_{n}(m_{0}) & = \dfrac{\sqrt{n} \ \overline{d}_{n}}{\widehat{\omega}_{n}}  
	\label{eq:eq7}
\end{align}
where  $\overline{d}_{n}=\sum_{t=k_{0}}^{T-h}\widehat{d}_{t}(m_{0})/n$ for
\begin{align}
\widehat{d}_{t}(m_{0}) = \hat{e}_{1,t+h|t}^{2}-\frac{1}{2}\left(
	\frac{n}{m_{0}} \ \hat{e}_{1,t+h|t}\hat{e}_{2,t+h|t} \ \textrm{I}(k_{0}\leq t\leq m_{0}+k_{0}-1)
\ + \ \right. \nonumber \\
	\left. \frac{n}{n-m_{0}} \ \hat{e}_{1,t+h|t}\hat{e}_{2,t+h|t} \ \textrm{I}(m_{0}+k_{0}<t\leq T-h)
	\right). \label{eq:eq8}
\end{align}

The test statistic in (\ref{eq:eq7}) is our main tool for conducting forecast encompassing tests across two nested models. Under the null hypothesis that model (\ref{eq:eq1}) forecast encompasses model (\ref{eq:eq2}), its numerator $\overline{d}_{n}$ which captures the covariance between $\hat{e}_{1,t+h|t}$ and 
($\hat{e}_{1,t+h|t}-\hat{e}_{2,t+h|t}$) will be small while under the alternative it will be positive/large so that the test is one-sided rejecting the null for sufficiently large and positive magnitudes of ${\cal E}_{n}(m_{0}) $.  \\
%Recalling that we take $m_{0}=[n \ \mu_{0}]$ in what follows it will be convenient to rescale the time axis 
%and write ${\cal E}_{n}(\mu_{0})$ for ${\cal E}_{n}([n\mu_{0}])$. \\

\noindent REMARK 2: To illustrate the intuition and benefits of the proposed statistic heuristically it is useful to obtain the asymptotic variance of $\sqrt{n} \ \overline{d}_{n}$ under the null hypothesis. 
Since under our nested environment and the null hypothesis the two models will share the same forecast errors asymptotically, let us replace $\hat{e}_{1,t+h|t}$ and $\hat{e}_{2,t+h|t}$ in $\overline{d}_{n}$ as defined in (\ref{eq:eq7})-(\ref{eq:eq8}) with $u_{t+h}$ and refer to this quantity as $\overline{d}_{n}^{0}$. Specifically, 
\begin{align}
\overline{d}_{n}^{0} & = \dfrac{\sum_{t=k_{0}}^{T-h}u_{t+h}^{2}}{n}-\dfrac{1}{2}\left(
\dfrac{\sum_{t=k_{0}}^{[n\mu_{0}]+k_{0}-1} u_{t+h}^{2}}{[n \mu_{0}]}
+\dfrac{\sum_{t=[n\mu_0]+k_{0}}^{T-h} u_{t+h}^{2}}{n-[n \mu_{0}]}\right). \label{eq:eq9}
\end{align}
\noindent Letting $\eta_{t+h}=u_{t+h}^{2}-E[u_{t+h}^{2}]$ with $E[u_{t+h}^{2}]$ assumed to be a positive constant we can equivalently write 
\begin{align}
	\overline{d}_{n}^{0} & = \dfrac{\sum_{t=k_{0}}^{T-h}\eta_{t+h}}{n}-\dfrac{1}{2}\left(
	\dfrac{\sum_{t=k_{0}}^{[n \mu_{0}]+k_{0}-1} \eta_{t+h}}{[n \mu_{0}]}
	+\dfrac{\sum_{t=[n\mu_{0}]+k_{0}}^{T-h} \eta_{t+h}}{n-[n \mu_{0}]}\right) \label{eq:eq10}
\end{align}
from which it is straightforward to establish that
\begin{align}
	\lim_{n \rightarrow \infty} V[\sqrt{n} \ \overline{d}_{n}^{0}] & = \dfrac{(1-2\mu_{0})^{2}}{4\mu_{0}(1-\mu_{0})} \ \sum_{j=-\infty}^{\infty} \gamma_{j}^{\eta}
	\label{eq:eq11}
\end{align}
\noindent assuming the quantity in the right hand side of (\ref{eq:eq11}) exists and is finite and where the $\gamma_{j}^{\eta}$'s refer to the autocovariances of $\eta_{t+h}$.  

The long-run variance in (\ref{eq:eq11}) illustrates two important points. Nested model encompassing tests that rely on our novel implementation of the population moment restriction in (\ref{eq:eq3}) are expected to have well-defined limiting distributions and will not suffer from variance degeneracy provided that $\mu_{0}$ is bounded away from $0$, $1$ and $1/2$. Importantly, these well-defined limiting distributions will hold under a $\sqrt{n}$-normalization of $\overline{d}_{n}^{0}$ .
The case $\mu_{0}=1/2$ reduces our setup to the traditional implementation of (\ref{eq:eq3}) via (\ref{eq:eq4}) and illustrates the impossibility of conducting valid inferences in such instances unless one wishes to forego Gaussian and nuisance parameter free asymptotics. 

%As we formalise further below 
%the practical implementation of the proposed test statistic will naturally require choosing a suitable magnitude %for $\mu_{0}$. Provided that $\mu_{0}\neq 1/2$ any choice of $\mu_{0} \in (0,1)$ 
%will ensure that under the null hypothesis ${\cal E}_{n}$ has a well-defined limit that does not depend on the %particular choice of $\mu_{0}$. Nevertheless, the chosen magnitude of $\mu_{0}$ may play an important role when %it comes to the power properties of our tests. One may expect that the closer 
%$\mu_{0}$ is to $1/2$ the stronger the power of the test will be. This is formally established in section 5 below %where we document the asymptotic power properties of the proposed test. 

\section{Assumptions and Null Asymptotics}

In this section we obtain the limiting distribution of our proposed test statistic under the null hypothesis of forecast encompassing as formulated in (\ref{eq:eq3}). We will find it convenient to operate under a set of high level assumptions that help emphasise the broad applicability of our results across a rich variety of settings.
Indeed, as under the null hypothesis the large sample behaviour of our proposed statistic is unaffected by whether predictors included in (\ref{eq:eq1})-(\ref{eq:eq2}) are stationary or highly persistent (e.g., having roots near unity in their AR representation) it becomes particularly convenient to operate under high level assumptions that compactly encompass such different scenarios. Our local power analysis further below introduces more primitive conditions highlighting specific scenarios under which our results hold. 
%These can accommodate for instance stationary as well as highly persistent processes in the set of predictors included in (\ref{eq:eq1})-(\ref{eq:eq2}). 

The null distribution of the proposed test statistic is established under the following set of assumptions. 

\vspace{0.2cm}
\noindent
{\bf Assumption A1.} (i) As $T\rightarrow \infty$, $\{T^{-1/2} \sum_{t=1}^{[Tr]}(u_{t}^{2}-E[u_{t}^{2}]), r\in [0,1]\} \stackrel{\cal D}\rightarrow  \{\phi \ W(r), r \in [0,1]\}$ in $\textrm{D}_{\mathds{R}}([0,1])$, where $\phi^{2} = \lim_{T\rightarrow \infty} V(\sqrt{T} \ \overline{\eta}_{t}) \in (0,\infty)$, $W(.)$ is a standard Brownian Motion and $\textrm{D}_{\mathds{R}}([0,1])$ is the space of cadlag functions on $[0,1]$. (ii) A consistent estimator of $\phi^{2}$ exists and is denoted $\widehat{\phi}^{2}$, that is ${\widehat{\phi}}^{2}\stackrel{p}\rightarrow \phi^{2}$ as $T\rightarrow \infty$. 

\noindent
{\bf Assumption A2.} For $i=1,2$ and as $T\rightarrow \infty$ we have
\begin{align}
\sup_{r \in [0,1]} \left|\frac{\sum_{t=k_{0}}^{k_{0}-h+[(T-k_{0})r]} \hat{e}_{i,t+h|t}^{2}}{\sqrt{n}}-\frac{\sum_{t=k_{0}}^{k_{0}-h+[(T-k_{0})r]} u_{t+h}^{2}}{\sqrt{n}}\right| & \stackrel{H_{0}}{=} o_{p}(1). \label{eq:eq12}
\end{align}

\noindent
{\bf Assumption A3.} The sample split fraction $\mu_{0}$ is bounded away from 0, $1/2$ and 1. 
%satisfies $\mu_{0}\in (0,1) \backslash \{\frac{1}{2}\}$

\vspace{0.2cm}
Assumptions {\bf A1}(i)-(ii) require the demeaned squared errors driving (\ref{eq:eq1})-(\ref{eq:eq2}) to satisfy 
a suitable functional central limit theorem with $\phi^{2}$ referring to the corresponding long-run variance.
Such a property is known to hold for a broad range of stochastic processes that are pertinent to our context
as when $u_{t}$ follows a finite or infinite order moving average process for instance (see e.g., \cite{solo1992}). Other examples of processes that satisfy {\bf A1}(i)-(ii) include a broad range of conditionally heteroskedastic processes under suitable finite moment requirements (see \cite{gkl2000, gkl2001}). 
%Examples of primitive conditions ensuring that {\bf A1}(i)-(ii) hold can be found in \cite{gkl2000} (Theorem 5.1), \cite{gkl2001} (Ex. 2.2 and Theorem 2.1). 
Assumption {\bf A2} requires the squared forecast errors to be consistent for their true counterparts. We may also note that {\bf A2} is assumed to hold for both $i=1$ and $i=2$ as we are operating under the null hypothesis. Assumption {\bf A2} can also be viewed as requiring the parameters of the fitted models to be estimable consistently so that asymptotically the $\hat{e}_{i,t}^{2}$'s  share the same stochastic properties as the $u_{t}^{2}$'s. For primitive conditions under which {\bf A2} holds 
we may consider the setting in \cite{dp2008} and \cite{ht2015} for strictly stationary and ergodic/mixing DGPs and \cite{bn2020} for environments that also allow near unit-root and long-memory processes. To clarify our use of the specific summation range in (\ref{eq:eq12}) it is useful to recall that h-steps ahead forecasts are built recursively starting from 
a given location $k_{0}$ so that we have a total of $n \equiv T-k_{0}-h+1$ forecast errors post-estimation. With the summand indexed as $t+h$ and taking $r=1$ for illustration purposes we have that $t+h$ ranges from $k_{0}+h$ to $T$ and the summation indices cover the entire set of available sample forecast errors. Finally, assumption {\bf A3} is crucial for the implementation of our sample-split based approach in nested settings. We may recall that the case $\mu_{0}=1/2$ is ruled out as it would result in $\sqrt{n} \overline{d}_{n}$ in (\ref{eq:eq7}) having a degenerate variance in the limit. 

Given assumptions {\bf A1}-{\bf A3} we can now proceed with the limiting behaviour of our proposed test statistic ${\cal E}_{n}(\mu_{0})$ in (\ref{eq:eq7}) under the null hypothesis. Before doing so it is important to 
focus on suitable candidates for the variance normalizer generically denoted as $\widehat{\omega}_{n}$ in (\ref{eq:eq7}). These will typically take the form of conventional HAC type formulations that are asymptotically equivalent albeit with potentially different finite sample distortions.  Here we opt to proceed analogously to how one would robustify a conventional t-ratio as given by (\ref{eq:eq7}) above. Letting $q_{t}(m_{0}) \equiv \hat{d}_{t}(m_{0})-\overline{d}_{n}(m_{0})$ with $\hat{d}_{t}(m_{0})$ as in (\ref{eq:eq8}), writing  $\hat{\gamma}^{q}_{\ell}(\ell)$ for the corresponding sample autocovariances of order $\ell$ of $q_{t}(m_{0})$, and using the Batlett kernel we take as our estimator $\hat{\omega}^{2}_{n}$ the quantity 
\begin{align}
\hat{\omega}_{n}^{2} & = \dfrac{1}{n}\sum_{t=1}^{n} q_{t}(m_{0})^{2}+\dfrac{2}{n}  \sum_{\ell=1}^{M} \left(1-\frac{\ell}{M}\right) \hat{\gamma}^{q}(\ell). \label{eq:eq13}
\end{align}
The above formulation via the Bartlett kernel ensures that $\hat{\omega}_{n}^{2}$ is non-negative while from \cite{a1991} a bandwidth satisfying $M=o(\sqrt{n})$ (e.g., $M=c \ n^{1/3}$ for some constant $c$) ensures consistency of $\hat{\omega}_{n}^{2}$. Given the expression of the limiting variance in (\ref{eq:eq11}) and Assumption {\bf A1(ii)} we will have $\hat{\omega}_{n}^{2} \stackrel{p}\rightarrow ((1-2\mu_{0})^{2}/4\mu_{0}(1-\mu_{0})) \phi^{2}$ as $n\rightarrow \infty$. 
The limiting distribution of ${\cal E}_{n}(\mu_{0})$ is now summarised in Proposition 1 below. 

\vspace{0.2cm}
\noindent
{\bf Proposition 1.} \emph{Under the null hypothesis, assumptions {\bf A1(i)-(ii)}, {\bf A2} and {\bf A3} and as $n\rightarrow \infty$
we have 
\begin{align}
	{\cal E}_{n}(\mu_{0}) & \stackrel{d}\rightarrow {\cal N}(0,1). \label{eq:eq14}
\end{align}
}
\noindent
The result in (\ref{eq:eq14}) illustrates how our proposed modification of the sample moment condition via (\ref{eq:eq5}) is able to conveniently bypass the variance degeneracy problem affecting existing inferences in nested settings. With the variance normaliser in (\ref{eq:eq13}) the limiting distribution of our proposed test statistic is free of nuisance parameter including the magnitude of $\mu_{0}$ used in its implementation. 

\section{Local Power Properties}

Our goal in this section is to evaluate the local power properties of ${\cal E}_{n}(\mu_{0})$ under departures from the null hypothesis. For this purpose we need to deviate from assumption {\bf A2} which was assumed to hold for both $i=1$ and $i=2$ as we now operate under model (\ref{eq:eq2}) parameterized as local to model (\ref{eq:eq1}). Here it will also be important to introduce more primitive conditions which allow us to explicitly distinguish the impact of having alternative predictor types on the power properties of our test (e.g., strictly stationary and ergodic versus highly persistent). We now assume that model (\ref{eq:eq2}) holds with ${\bm \gamma}={\bm c}/T^{\kappa}$ for ${\bm c}\neq 0$ and some suitable $\kappa$ and we explore the limiting behaviour of ${\cal E}_{n}(\mu_{0})$ under such a local to the null alternative. Although the null asymptotics remained unaffected by whether predictors were purely stationary or persistent this will no longer be the case for the power properties of our test. In what follows we distinguish between two alternative scenarios for the stochastic properties of the predictors in (\ref{eq:eq2}). The first scenario we consider is one where predictors satisfy a standard law of large numbers as in {\bf B2} below while in the second scenario ({\bf C2}) we model them as persistent process via an autoregressive process with roots close to unity. In what follows it will be useful to recall that the vector $\widetilde{\bm x}_{12,t}$ combines and stacks the intercept, the ${\bm x}_{1,t}$'s and ${\bm x}_{2,t}$'s so that $\widetilde{\bm x}_{12,t}=(\widetilde{\bm x}_{1,t}^{\top},{\bm x}_{2,t}^{\top})^{\top}$ where $\widetilde{\bm x}_{1,t}=(1,{\bm x}_{1,t}^{\top})^{\top}$. \\

\noindent
{\bf Assumption B1.} Assumptions {\bf A1}, {\bf A2}(ii) and {\bf A3} hold. 

\noindent
{\bf Assumption B2.} (i) As $T\rightarrow \infty$ it holds that $\sup_{\lambda \in [0,1]}\|\sum_{t=1}^{[T\lambda]} \widetilde{{\bm x}}_{12,t-h}\widetilde{{\bm x}}_{12,t-h}^{\top}/T-\lambda {\bm Q}\|=o_{p}(1)$ for some symmetric positive definite deterministic matrix ${\bm Q}$. (ii) A functional central limit theorem holds for $\{ \widetilde{\bm x}_{12,t} w_{t+h}\}$. That is, $\sum_{t=1}^{[T\lambda]}\widetilde{\bm x}_{12,t-h}w_{t}/\sqrt{T}\stackrel{\cal D}\rightarrow {\bm \Omega}_{\infty}^{1/2} {\bm W}(\lambda)$ with ${\bm W}(\lambda)$ denoting a $p_{1}+p_{2}+1$ dimensional standard Brownian Motion and ${\bm \Omega}_{\infty}>0$ the long run variance of $\{\widetilde{\bm x}_{12,t-h}w_{t}\}$.

\vspace{0.2cm}
The current setup based on assumptions {\bf B1-B2} corresponds to en environment with strictly stationary and ergodic predictors similar to the setup in Hansen and Timmermann (2015). Note that the explicit formulation of the FCLT in {\bf B2}(ii) is not strictly necessary in the sense that we only require the sample moment to be stochastically bounded which is in turn ensured by the FCLT. Finally, we also partition ${\bm Q}$ as 
${\bm Q}_{ij}$ for $i,j=1,2$  along the two components of $\widetilde{\bm x}_{12,t}$. Our first result is summarised in Proposition 2.  

\vspace{0.2cm}
\noindent
{\bf Proposition 2.} \emph{Suppose model (\ref{eq:eq2}) holds with ${\bm \gamma}={\bm c}/T^{1/4}$ for ${\bm c} \neq 0$. Under assumptions {\bf B1} and {\bf B2} and as $n\rightarrow \infty$ we have
	\begin{align}
		{\cal E}_{n}(\mu_{0}) & \stackrel{d}\rightarrow {\cal N}(0,1)
		+\sqrt{1-\pi_{0}} \left(\sqrt{\dfrac{4\mu_{0}(1-\mu_{0})}{(1-2\mu_{0})^{2}\phi^{2}}} \ {\bm c}^\top ({\bm Q}_{22}-{\bm Q}_{21}{\bm Q}_{11}^{-1}{\bm Q}_{12}) {\bm c}\right)
		. \label{eq:eq15}  
	\end{align}
}

The formulation in (\ref{eq:eq15}) is particularly helpful for understanding the influence of $\mu_{0}$ on the power properties of the proposed test. All other things being equal we can infer for instance that magnitudes of $\mu_{0}$ in the vicinity of $1/2$ will be associated with more favourable test power. Similarly, power will also be influenced by the variance of the predictors in the larger model and by how these predictors correlate with the predictors included in the smaller model, as captured by the quadratic form ${\bm c}^\top ({\bm Q}_{22}-{\bm Q}_{21}{\bm Q}_{11}^{-1}{\bm Q}_{12}) \bm c$. All other things being equal, we should expect more favourable power properties when the ${\bm x}_{2,t}$'s are more persistent. This is formalized in the next scenario 
which illustrates the ability of the proposed test to detect departures that are nearer the null hypothesis when predictors are persistent. 

The literature on modelling persistence offers a broad range of alternative parameterizations designed to capture strong dependence across time. As our goal is to document the role of persistence on local power the specific nature of the type of persistence is not of particular importance. Here we opt to capture persistence via a mildly integrated environment by letting the predictors driving (\ref{eq:eq2}) follow a simple mildly integrated VAR. Letting $\bm x_{12,t}=({\bm x}_{1,t}^{\top},{\bm x}_{2,t}^{\top})^{\top}$ (as opposed to $\widetilde{x}_{12,t}$ defined earlier) we consider the following specification
\begin{align}
{\bm x}_{12,t} & = \left(I_{p_{1}+p_{2}}-\dfrac{\cal B}{T^{\alpha}}\right){\bm x}_{12,t-1}+{\bm v}_{t} \label{eq:eq16}
\end{align} 
\noindent
where ${\cal B}=diag({b}_{1},\ldots,b_{p_{1}+p_{2}})$ with $b_{i}<0$, $\alpha \in (0,1)$ and ${\bm v}_{t}$ is a random disturbance vector. We note from (\ref{eq:eq16}) that as $\alpha$ approaches 1 the components of 
$\bm x_{12,t}$ become increasingly persistent. To proceed with a set of high level assumptions that 
can accommodate such dynamics we introduce the following normalization matrix designed to suitably normalize each individual
component of $\widetilde{\bm x}_{12,t}=(1,{\bm x}_{12,t}^{\top})^{\top}$ and given by
\begin{align}
{\mathds{D}}_{T} & = 
\begin{pmatrix}
\frac{1}{\sqrt{T}} & 0 & 0 \\
0 & \frac{1}{T^{\frac{1+\alpha}{2}}} & 0 \\
0 & 0 & \frac{1}{T^{\frac{1+\alpha}{2}}}	
\end{pmatrix}
.
\label{eq:eq17}
\end{align}
\noindent
We now replace Assumptions ${\bf B1}$ and ${\bf B2}$ above with 

\noindent
{\bf Assumption C1.} Assumptions {\bf A1}, {\bf A2}(ii) and {\bf A3} hold. 

\noindent
{\bf Assumption C2.} (i) As $T\rightarrow \infty$ it holds that $\sup_{\lambda}\|\mathds{D}_{T} \sum_{t=1}^{[T\lambda]} \widetilde{{\bm x}}_{12,t-h}\widetilde{{\bm x}}_{12,t-h}'\mathds{D}_{T} \stackrel{p} \rightarrow \lambda {\bm V}^{b}\|=o_{p}(1)$ for a non-random symmetric positive definite ${\bm V}^{b}$. (ii) 
 $\sup_{\lambda \in [0,1]}\|\mathds{D}_{T}\sum_{t=1}^{[T\lambda]} \widetilde{{\bm x}}_{12,t-h}u_{t+h}\|=O_{p}(1)$.

The properties of mildly integrated processes as in (\ref{eq:eq16}) and leading to properties such as {\bf C2} above have been extensively studied both in univariate and multivariate settings (see \cite{mp2009}, \cite{pm2009}, \cite{kms2015}, \cite{bd2015}, \cite{pcb2023} amongst others). The typical elements of ${\bm V}^{b}$ for instance are known to depend on the $b_{i}$ terms 
in ${\cal B}$ as well as the variance and covariances of the components of ${\bm v}_{t}$. The following Proposition summarizes the local power properties of ${\cal E}_{n}$ in a setting with mildly integrated predictors. \\

\noindent
{\bf Proposition 3.} \emph{Suppose model (\ref{eq:eq2}) holds with ${\bm \gamma}={\bm c}/T^{\frac{1}{4}+\frac{\alpha}{2}}$ for ${\bm c} \neq 0$. Under assumptions {\bf C1} and {\bf C2} and as $T\rightarrow \infty$ we have
	\begin{align}
		{\cal E}_{n}(\mu_{0}) & \stackrel{d}\rightarrow {\cal N}(0,1)
		+\sqrt{1-\pi_{0}} \left(\sqrt{\dfrac{4\mu_{0}(1-\mu_{0})}{(1-2\mu_{0})^{2}\phi^{2}}} \ {\bm c}^\top 
		({\bm V}_{22}^{b}-{\bm V}_{21}^{b}({\bm V^{b}}_{11})^{-1}{\bm V}^{b}_{12})
		 {\bm c}\right)
		. \label{eq:eq18}
	\end{align}
}

Comparing the above two scenarios we first note that the presence of persistent additional predictors in the larger model ensures more favourable power behaviour in the sense that ${\bm \gamma}$ is allowed to vanish faster. This is clearly captured by the normalization $T^{\frac{1}{4}+\frac{\alpha}{2}}$ compared with $T^{\frac{1}{4}}$ so that persistence captured by a positive $\alpha \in (0,1)$ results in important differences in the ability of the test statistic to detect local departures from the null. 
Secondly, we also note that the role of $\mu_{0}$ is unaffected by whether we are under a persistent or standard setting. In both instances a magnitude in the vicinity of $1/2$ is expected to ensure more favourable empirical powers.

\section{Empirical Size and Power Properties}

In this section we illustrate the behaviour of our proposed test statistic in finite samples. We initially focus on its empirical size properties and subsequently investigate its ability to detect departures from the null hypothesis. Recalling that the hypothesis test of interest is one-sided to the right we implement all our size and power analyses using a 10\% nominal level.  Our size experiments use samples of size $T=250, 500, 1000$ and we set $\pi_{0}=0.25$ so that with $h=1$ for instance the effective sample sizes are 187, 375 and 750 respectively. For our power experiments we fix the sample size at $T=500$ and evaluate the ability of our statistics to detect progressive departures from the null by increasing the true values of slope parameters. All our experiments consider both single step and multi-steps ahead forecasts with $h=1,4,12, 24$. Outcomes are obtained as averages across 10000 replications. 

\subsection{Empirical Size}

\noindent
{\bf \small DGP1}: The first DGP consists of the following predictive regression model with two predictors 
\begin{align}
y_{t+h} & = \beta_{1} y_{t}+ \beta_{2} x_{t}+w_{t+h} \nonumber \\
x_{t} & = \rho x_{t-1}+v_{t} \nonumber \\
w_{t+h} & = \sum_{j=0}^{h-1}\theta^{j} \epsilon_{t+h-j} \label{eq:eq19}
\end{align}
\noindent
where $(\epsilon_{t},v_{t})^\top \sim \textrm{NID}({\bm 0},\bm \Sigma)$. We set $\beta_{1}=0.3$ and $\beta_{2}=0$ throughout so that the benchmark model is given by an AR(1) process while the larger model has a single additional predictor. Note also that $w_{t+h}$ is parameterized as an MA(h-1) process for which we set $\theta=0.5$. We consider $\rho \in \{0.25, 0.90, 0.95\}$ so that our experiments cover both quickly mean reverting as well as highly persistent predictors. Finally we consider two alternative scenarios for ${\bm \Sigma} \in \{\bm \Sigma_{1},\bm \Sigma_{2}\}$ with $\bm \Sigma_{1}=\{\{1,0\},\{0,0.25\}\}$ 
 and $\bm \Sigma_{2}=\{\{1,-0.4\},\{-0.4,0.25\}\}$. This latter scenario allows for a strong negative correlation between the 
 $\epsilon_{t}'s$ and $v_{t}'s$ (i.e., correlation of -0.80). At this stage it is also useful to recall that under the null hypothesis forecasts from the larger model are encompassed in the forecasts from the benchmark model so that the population restriction in (\ref{eq:eq3}) holds when $\beta_{2}=0$. 
 
 % Table generated by Excel2LaTeX from sheet 'dgp1a-d-alrv'
 \begin{table}[htbp]
 	\centering
 \tabcolsep=0.16cm
 \caption{Empirical Size of ${\cal E}_{n}(\mu_{0})$ - $\textrm{\bf DGP1}(\bm \Sigma_{1})$}
 \vspace{0.4cm}
 \scalebox{0.88}{\begin{tabular}{lcccccccccccc} \hline 
 		& \multicolumn{4}{c}{$\rho=0.25$} & \multicolumn{4}{c}{$\rho=0.90$} & \multicolumn{4}{c}{$\rho=0.95$} \\ \hline
 		$\mu_{0}$ & 0.30  & 0.35  & 0.40  & 0.45  & 0.30  & 0.35  & 0.40  & 0.45  & 0.30  & 0.35  & 0.40  & 0.45 \\ \hline
 		& \multicolumn{12}{c}{$h=1$} \\ 
 		T=250 & 0.116 & 0.107 & 0.103 & 0.092 & 0.121 & 0.111 & 0.107 & 0.091 & 0.125 & 0.115 & 0.107 & 0.093 \\
 		T=500 & 0.118 & 0.112 & 0.108 & 0.102 & 0.110 & 0.108 & 0.106 & 0.098 & 0.116 & 0.110 & 0.105 & 0.099 \\
 		T=1000 & 0.113 & 0.105 & 0.103 & 0.095 & 0.106 & 0.105 & 0.099 & 0.098 & 0.103 & 0.103 & 0.097 & 0.094 \\
 		& \multicolumn{12}{c}{$h=4$} \\
 		T=250 & 0.137 & 0.128 & 0.119 & 0.105 & 0.138 & 0.129 & 0.119 & 0.100 & 0.138 & 0.132 & 0.120 & 0.100 \\
 		T=500 & 0.129 & 0.126 & 0.117 & 0.109 & 0.130 & 0.124 & 0.115 & 0.098 & 0.135 & 0.130 & 0.120 & 0.103 \\
 		T=1000 & 0.119 & 0.114 & 0.108 & 0.106 & 0.121 & 0.115 & 0.108 & 0.097 & 0.122 & 0.115 & 0.107 & 0.098 \\
 		& \multicolumn{12}{c}{$h=12$} \\
 		T=250 & 0.138 & 0.128 & 0.123 & 0.110 & 0.144 & 0.134 & 0.128 & 0.120 & 0.148 & 0.132 & 0.126 & 0.109 \\
 		T=500 & 0.131 & 0.125 & 0.123 & 0.111 & 0.130 & 0.124 & 0.117 & 0.102 & 0.132 & 0.127 & 0.120 & 0.106 \\
 		T=1000 & 0.124 & 0.122 & 0.117 & 0.110 & 0.119 & 0.115 & 0.108 & 0.107 & 0.123 & 0.121 & 0.114 & 0.107 \\
 		& \multicolumn{12}{c}{$h=24$} \\
 		T=250 & 0.131 & 0.125 & 0.117 & 0.107 & 0.140 & 0.133 & 0.131 & 0.121 & 0.146 & 0.138 & 0.132 & 0.123 \\
 		T=500 & 0.130 & 0.124 & 0.115 & 0.109 & 0.131 & 0.122 & 0.115 & 0.109 & 0.131 & 0.127 & 0.124 & 0.113 \\
 		T=1000 & 0.120 & 0.118 & 0.113 & 0.108 & 0.125 & 0.118 & 0.110 & 0.107 & 0.126 & 0.120 & 0.111 & 0.098 \\
 	\end{tabular}
}
 	\label{tab:addlabel}
 \end{table}%

 Empirical size outcomes associated with this first DGP are presented in Table 1 for the case $\bm \Sigma=\bm \Sigma_{1}$ and in Table 2 for for the case $\bm \Sigma=\bm \Sigma_{2}$. These clearly support our asymptotic theory whereby ${\cal E}_{n}(\mu_{0})\stackrel{d}\rightarrow {\cal N}(0,1)$ with the asymptotic approximation seen to provide accurate size estimates across virtually all scenarios. (i) A first observation to make is the robustness of outcomes to the specific choice of the sample split fraction $\mu_{0}$ used in the implementation of our test statistic, as suggested by the asymptotic theory of Proposition 1. (ii) A second important observation to make is the overall efficacy of the chosen Newey-West type long run variance normalizer in cleaning out the dependence induced by the MA(h-1) errors for $h>1$. (iii) Finally, the third important point to infer from these outcomes is the robustness of outcomes to the degree of persistence of the predictor $x_{t}$ as also expected from the underlying asymptotic theory. Indeed, we may note the strong similarities in the empirical sizes across $\rho=0.25$, $\rho=0.90$ and $\rho=0.95$.  
 
 % Table generated by Excel2LaTeX from sheet 'dgp1a-d-alrv'
 \begin{table}[htbp]
 	\centering
 \tabcolsep=0.16cm
 \caption{Empirical Size of ${\cal E}_{n}(\mu_{0})$ - $\textrm{\bf DGP1}(\bm \Sigma_{2})$}
 \vspace{0.4cm}
 \scalebox{0.88}{	\begin{tabular}{lcccccccccccc} \hline
 		& \multicolumn{4}{c}{$\rho=0.25$} & \multicolumn{4}{c}{$\rho=0.90$} & \multicolumn{4}{c}{$\rho=0.95$} \\ \hline
 		$\mu_{0}$ & 0.30  & 0.35  & 0.40  & 0.45  & 0.30  & 0.35  & 0.40  & 0.45  & 0.30  & 0.35  & 0.40  & 0.45 \\ \hline
 		& \multicolumn{12}{c}{$h=1$} \\
 		T=250 & 0.108 & 0.100 & 0.099 & 0.090 & 0.117 & 0.114 & 0.107 & 0.092 & 0.124 & 0.113 & 0.104 & 0.096 \\
 		T=500 & 0.108 & 0.102 & 0.098 & 0.089 & 0.114 & 0.106 & 0.104 & 0.097 & 0.111 & 0.105 & 0.102 & 0.098 \\
 		T=1000 & 0.111 & 0.107 & 0.105 & 0.099 & 0.105 & 0.105 & 0.102 & 0.096 & 0.112 & 0.111 & 0.108 & 0.103 \\
 		& \multicolumn{12}{c}{$h=4$} \\
 		T=250 & 0.143 & 0.134 & 0.129 & 0.112 & 0.134 & 0.127 & 0.117 & 0.099 & 0.136 & 0.127 & 0.117 & 0.100 \\
 		T=500 & 0.135 & 0.123 & 0.118 & 0.109 & 0.132 & 0.118 & 0.112 & 0.103 & 0.132 & 0.130 & 0.120 & 0.099 \\
 		T=1000 & 0.128 & 0.125 & 0.116 & 0.111 & 0.119 & 0.112 & 0.109 & 0.099 & 0.120 & 0.120 & 0.113 & 0.103 \\
 		& \multicolumn{12}{c}{$h=12$} \\
 		T=250 & 0.134 & 0.126 & 0.115 & 0.105 & 0.136 & 0.127 & 0.119 & 0.105 & 0.139 & 0.131 & 0.117 & 0.106 \\
 		T=500 & 0.132 & 0.123 & 0.116 & 0.107 & 0.131 & 0.123 & 0.115 & 0.103 & 0.132 & 0.122 & 0.116 & 0.102 \\
 		T=1000 & 0.120 & 0.114 & 0.112 & 0.105 & 0.120 & 0.114 & 0.106 & 0.097 & 0.126 & 0.122 & 0.115 & 0.103 \\
 		& \multicolumn{12}{c}{$h=24$} \\
 		T=250 & 0.144 & 0.138 & 0.127 & 0.117 & 0.142 & 0.134 & 0.124 & 0.114 & 0.141 & 0.133 & 0.127 & 0.117 \\
 		T=500 & 0.126 & 0.119 & 0.108 & 0.103 & 0.129 & 0.125 & 0.120 & 0.104 & 0.129 & 0.126 & 0.120 & 0.103 \\
 		T=1000 & 0.121 & 0.121 & 0.110 & 0.106 & 0.124 & 0.115 & 0.113 & 0.097 & 0.122 & 0.117 & 0.113 & 0.107 \\
 	\end{tabular}
}
 	\label{tab:addlabel}%
 \end{table}%

These outcomes also point to some weaknesses of our proposed test statistic when a small sample size is combined with a persistent predictor {\it and} longer-horizon based inferences. The simultaneous combination of a small $T$, $\rho$ near unity and $h$ large (e.g., $h=24$) can be seen to be associated with inflated sizes, albeit quite mildly. Such distortions are not unique to our setting and are commonly encountered in the broader testing literature when one considers Newey-West type standard errors in environments characterised by moderate to strong dependence in time-series. Furthermore, we may also observe that for $\mu_{0}$ in the vicinity of 0.5 distortions are very mild under $T=250$ while empirical and nominal sizes match perfectly for $T>250$ even under the most 
challenging of scenarios (e.g., $h=24$).    

The outcomes in Table 2 are both qualitatively and quantitatively very similar to those of Table 1. The only difference between the two scenarios is in the parameterization of $\bm \Sigma= {\bm \Sigma_{2}}$ which allows for a fairly strong negative correlation between the shocks to $y_{t}$ and the shocks to the predictor $x_{t}$. Such a strong negative correlation is a common occurrence in predictive regressions that link stock returns to fundamentals such as dividend yields or price-to-earnings ratios. Our proposed method is clearly able to display favourable properties in such environments. \\

\noindent
{\bf DGP2}: Our second DGP is designed to illustrate a potentially useful application of our method in data rich environments. The setting we consider is that of a factor-augmented regression with the objective of 
implementing a forecast encompassing test between an autoregressive specification augmented with a diffusion index type single composite factor and one that excludes it. Specifically, we consider
\begin{align}
y_{t+h} & = \alpha + \beta_{1} y_{t}+ \beta_{2}f_{t} + w_{t+h} \nonumber \\
w_{t+h} & = \sum_{j=1}^{h-1} \theta^{j} v_{t+h-j} \label{eq:eq20}
\end{align}
and set $\beta_{2}=0$ for evaluating size properties. Note that as in the previous DGP we have added moving average dynamics to the error process involving a horizon greater than one. The fitted specification from which forecasts are built is given by (\ref{eq:eq20}) with $f_{t}$ replaced by $\hat{f}_{t}$. For this purpose we operate with the following approximate factor model
\begin{align}
X_{it} & = \lambda_{i} f_{t}+e_{it} \nonumber \\
f_{t} & = \alpha_{1} f_{t-1}+u_{t} \nonumber \\
e_{it} & = \rho_{i} e_{it-1}+\epsilon_{it}  \label{eq:eq21}
\end{align}
from which $\hat{f}_{t}$ is estimated as in \cite{bn2002}. Note that in this experiment we set aside the uncertainty that arises when estimating the number of factors by assuming that the latter are known to be equal to one. This is reasonable for our purpose as we wish to isolate the properties of our proposed encompassing test without blurring its properties via additional estimation and specification uncertainty beyond that involved in using $\hat{f}_{t}$ as a proxy for the unobserved $f_{t}$. It is also implicitly understood that the properties of our test remain unaffected by the inclusion of a generated regressor. In the present context this is supported by the findings in \cite{gmp2017} where the authors show that the estimation of factors does not affect the asymptotics of predictive accuracy comparisons. 

Results for this experiment are presented in Table 3 across two scenarios for the cross-sectional and time series dimensions $(N,T)\in\{(100,250),(500,500)\}$. Outcomes are very much in line with our earlier observations for {\bf DGP1}. We note for instance that for $\mu_{0}$ in the close vicinity of 0.5, empirical sizes almost perfectly match their nominal counterpart of 10\%. Some finite sample size distortions do tend to occur under small sample sizes and magnitudes of $\mu_{0}$ such as 0.3. As we discuss further below and as established in our local power analysis this behaviour of ${\cal E}_{n}(\mu_{0})$ when $\mu_{0}$ is in the vicinity of 0.5 is highly encouraging as power has been shown to peak for such magnitudes while from Table 2 we can also observe that in finite samples the size behaviour of our test is even more favourable for such magnitudes of $\mu_{0}$.

% Table generated by Excel2LaTeX from sheet 'dgp2-d-T10alrv'
\begin{table}[htbp]
		\centering
	\tabcolsep=0.16cm
	\caption{Empirical Size of ${\cal E}_{n}(\mu_{0})$ - $\textrm{\bf DGP2}$}
	\vspace{0.4cm}
	\begin{tabular}{lcccc} \hline
		$\mu_{0}$ & 0.30  & 0.35  & 0.40  & 0.45 \\ \hline
		& \multicolumn{4}{c}{(N,T)=(100,250)} \\
		h=1   & 0.120 & 0.117 & 0.107 & 0.097 \\
		h=4   & 0.138 & 0.131 & 0.117 & 0.101 \\
		h=12  & 0.139 & 0.132 & 0.120 & 0.111 \\
		h=24  & 0.141 & 0.130 & 0.119 & 0.107 \\ \hline
		& \multicolumn{4}{c}{(N,T)=(500,500)} \\
		h=1   & 0.107 & 0.102 & 0.102 & 0.096 \\
		h=4   & 0.132 & 0.127 & 0.118 & 0.110 \\
		h=12  & 0.134 & 0.122 & 0.115 & 0.107 \\
		h=24  & 0.130 & 0.124 & 0.116 & 0.105 \\
	\end{tabular}%
	\label{tab:addlabel}%
\end{table}%

 \subsection{Empirical Power}
 
 \noindent
 {\bf DGP1}: To explore power in the context of the DGP in (\ref{eq:eq17}) we continue to set $\beta_{1}=0.3$ as above and now 
 increase the magnitude $\beta_{2}$ away from zero incrementally, given a fixed sample size which we set at $T=500$. Specifically, we consider  
 \begin{center}
 $\beta_{2}\in \{0.10;0.20;0.30;0.35;0.40;0.45;0.50;0.55;0.60\}$. 
 \end{center}
and outcomes for $\bf DGP1(\Sigma_{1})$ are presented in Table 4 below. As expected from our theoretical local power analysis we can clearly observe the correct decision frequencies converging to 100\% as $\beta_{2}$ moves away from 0. Key takeaway points from these outcomes are as follows: (i) power improves as $\mu_{0}$ is selected nearer its boundary of 0.5, (ii) power is substantially higher for persistent predictors (e.g., $\rho=0.90$ or $\rho=0.95$ vs $\rho=0.25$), (iii) there appears to be a meaningful 
decline in power as we move away from single period forecasts with the latter exhibiting more favourable detection frequencies. Interestingly, for $h>1$, differences across $h$ are either very mild or negligible. \\

\noindent
{\bf DGP2}: Power outcomes for {\bf DGP2} are presented in Table 5 and are broadly aligned with the above points. Under $(N,T)=(500,500)$ for instance which is often the norm in many empirical applications, we can note the very rapid convergence of the correct detection frequencies to 100\% (for $\mu_{0}$ in the vicinity of 0.5 in particular). Power is also at its best for single period forecasts while dropping markedly as $h$ is increased to $h=4$ or beyond. Interestingly, there appears to be only marginal differences in power across $h$ when excluding the case $h=1$.

% Table generated by Excel2LaTeX from sheet 'dgp1a-T10alrvd-power'
\begin{table}[htbp]
 	\centering
\tabcolsep=0.16cm
\caption{EmpiricalPower of ${\cal E}_{n}(\mu_{0})$ - $\textrm{\bf DGP1}(\bm \Sigma_{1})$}
\vspace{0.4cm}
\scalebox{0.88}{\begin{tabular}{lcccccccccccc} \hline
		$\mu_{0}$ & 0.30  & 0.35  & 0.40  & 0.45  & 0.30  & 0.35  & 0.40  & 0.45  & 0.30  & 0.35  & 0.40  & 0.45 \\ \hline
		& \multicolumn{4}{c}{$\rho=0.25$} & \multicolumn{4}{c}{$\rho=0.90$} & \multicolumn{4}{c}{$\rho=0.95$} \\ \hline
		& \multicolumn{12}{c}{$h=1$} \\
		$\beta_{2}=0.10$ & 0.126 & 0.128 & 0.136 & 0.162 & 0.201 & 0.236 & 0.313 & 0.455 & 0.308 & 0.384 & 0.507 & 0.701 \\
		$\beta_{2}=0.20$ & 0.189 & 0.210 & 0.272 & 0.409 & 0.544 & 0.677 & 0.833 & 0.953 & 0.797 & 0.901 & 0.967 & 0.995 \\
		$\beta_{2}=0.30$ & 0.305 & 0.380 & 0.518 & 0.728 & 0.887 & 0.961 & 0.993 & 0.999 & 0.982 & 0.996 & 1.000 & 1.000 \\
		$\beta_{2}=0.35$ & 0.378 & 0.485 & 0.649 & 0.848 & 0.957 & 0.991 & 0.999 & 1.000 & 0.996 & 1.000 & 1.000 & 1.000 \\
		$\beta_{2}=0.40$ & 0.479 & 0.607 & 0.771 & 0.931 & 0.984 & 0.998 & 1.000 & 1.000 & 1.000 & 1.000 & 1.000 & 1.000 \\
		$\beta_{2}=0.45$ & 0.588 & 0.727 & 0.873 & 0.974 & 0.996 & 0.999 & 1.000 & 1.000 & 1.000 & 1.000 & 1.000 & 1.000 \\
		$\beta_{2}=0.50$ & 0.690 & 0.824 & 0.938 & 0.990 & 0.999 & 1.000 & 1.000 & 1.000 & 1.000 & 1.000 & 1.000 & 1.000 \\
		$\beta_{2}=0.55$ & 0.781 & 0.896 & 0.972 & 0.997 & 1.000 & 1.000 & 1.000 & 1.000 & 1.000 & 1.000 & 1.000 & 1.000 \\
		$\beta_{2}=0.60$ & 0.852 & 0.944 & 0.991 & 1.000 & 1.000 & 1.000 & 1.000 & 1.000 & 1.000 & 1.000 & 1.000 & 1.000 \\
		& \multicolumn{12}{c}{$h=4$} \\
		$\beta_{2}=0.10$ & 0.139 & 0.135 & 0.139 & 0.151 & 0.183 & 0.199 & 0.224 & 0.271 & 0.249 & 0.276 & 0.325 & 0.400 \\
		$\beta_{2}=0.20$ & 0.181 & 0.190 & 0.215 & 0.286 & 0.375 & 0.445 & 0.549 & 0.678 & 0.567 & 0.664 & 0.772 & 0.873 \\
		$\beta_{2}=0.30$ & 0.234 & 0.271 & 0.349 & 0.503 & 0.649 & 0.755 & 0.859 & 0.937 & 0.865 & 0.929 & 0.969 & 0.990 \\
		$\beta_{2}=0.35$ & 0.282 & 0.339 & 0.447 & 0.637 & 0.770 & 0.866 & 0.938 & 0.979 & 0.931 & 0.973 & 0.992 & 0.999 \\
		$\beta_{2}=0.40$ & 0.340 & 0.416 & 0.541 & 0.749 & 0.866 & 0.935 & 0.978 & 0.995 & 0.972 & 0.992 & 0.998 & 1.000 \\
		$\beta_{2}=0.45$ & 0.404 & 0.504 & 0.656 & 0.846 & 0.935 & 0.977 & 0.993 & 0.999 & 0.991 & 0.998 & 0.999 & 1.000 \\
		$\beta_{2}=0.50$ & 0.480 & 0.591 & 0.748 & 0.910 & 0.970 & 0.993 & 0.998 & 1.000 & 0.996 & 0.999 & 1.000 & 1.000 \\
		$\beta_{2}=0.55$ & 0.539 & 0.668 & 0.818 & 0.950 & 0.988 & 0.997 & 0.999 & 1.000 & 0.999 & 1.000 & 1.000 & 1.000 \\
		$\beta_{2}=0.60$ & 0.618 & 0.749 & 0.886 & 0.977 & 0.994 & 0.999 & 1.000 & 1.000 & 1.000 & 1.000 & 1.000 & 1.000 \\
		& \multicolumn{12}{c}{$h=12$} \\
		$\beta_{2}=0.10$ & 0.141 & 0.140 & 0.137 & 0.148 & 0.189 & 0.199 & 0.221 & 0.261 & 0.243 & 0.271 & 0.310 & 0.378 \\
		$\beta_{2}=0.20$ & 0.177 & 0.192 & 0.209 & 0.281 & 0.374 & 0.439 & 0.534 & 0.647 & 0.555 & 0.648 & 0.746 & 0.841 \\
		$\beta_{2}=0.30$ & 0.238 & 0.277 & 0.352 & 0.504 & 0.642 & 0.740 & 0.838 & 0.913 & 0.864 & 0.927 & 0.967 & 0.987 \\
		$\beta_{2}=0.35$ & 0.284 & 0.335 & 0.438 & 0.627 & 0.769 & 0.854 & 0.927 & 0.972 & 0.931 & 0.968 & 0.987 & 0.997 \\
		$\beta_{2}=0.40$ & 0.326 & 0.397 & 0.526 & 0.732 & 0.867 & 0.935 & 0.973 & 0.992 & 0.976 & 0.991 & 0.998 & 0.999 \\
		$\beta_{2}=0.45$ & 0.391 & 0.483 & 0.624 & 0.826 & 0.925 & 0.967 & 0.990 & 0.997 & 0.993 & 0.999 & 1.000 & 1.000 \\
		$\beta_{2}=0.50$ & 0.454 & 0.561 & 0.723 & 0.892 & 0.968 & 0.990 & 0.998 & 1.000 & 0.996 & 0.999 & 1.000 & 1.000 \\
		$\beta_{2}=0.55$ & 0.529 & 0.650 & 0.801 & 0.941 & 0.987 & 0.996 & 0.999 & 1.000 & 0.999 & 1.000 & 1.000 & 1.000 \\
		$\beta_{2}=0.60$ & 0.598 & 0.731 & 0.872 & 0.969 & 0.994 & 0.999 & 1.000 & 1.000 & 1.000 & 1.000 & 1.000 & 1.000 \\
		& \multicolumn{12}{c}{$h=24$} \\
		$\beta_{2}=0.10$ & 0.135 & 0.136 & 0.134 & 0.146 & 0.191 & 0.206 & 0.230 & 0.266 & 0.254 & 0.282 & 0.326 & 0.388 \\
		$\beta_{2}=0.20$ & 0.163 & 0.175 & 0.206 & 0.279 & 0.358 & 0.427 & 0.519 & 0.626 & 0.554 & 0.637 & 0.732 & 0.822 \\
		$\beta_{2}=0.30$ & 0.233 & 0.266 & 0.344 & 0.485 & 0.631 & 0.726 & 0.823 & 0.907 & 0.869 & 0.925 & 0.964 & 0.985 \\
		$\beta_{2}=0.35$ & 0.271 & 0.328 & 0.428 & 0.606 & 0.763 & 0.850 & 0.923 & 0.966 & 0.930 & 0.966 & 0.986 & 0.995 \\
		$\beta_{2}=0.40$ & 0.330 & 0.396 & 0.515 & 0.709 & 0.849 & 0.920 & 0.967 & 0.990 & 0.969 & 0.990 & 0.997 & 0.999 \\
		$\beta_{2}=0.45$ & 0.374 & 0.468 & 0.608 & 0.802 & 0.918 & 0.963 & 0.987 & 0.996 & 0.993 & 0.998 & 1.000 & 1.000 \\
		$\beta_{2}=0.50$ & 0.442 & 0.553 & 0.711 & 0.887 & 0.958 & 0.984 & 0.996 & 0.999 & 0.997 & 1.000 & 1.000 & 1.000 \\
		$\beta_{2}=0.55$ & 0.515 & 0.634 & 0.791 & 0.932 & 0.980 & 0.996 & 0.999 & 1.000 & 0.998 & 1.000 & 1.000 & 1.000 \\
		$\beta_{2}=0.60$ & 0.582 & 0.711 & 0.859 & 0.966 & 0.991 & 0.997 & 1.000 & 1.000 & 1.000 & 1.000 & 1.000 & 1.000 \\
	\end{tabular}}
	\label{tab:addlabel}%
\end{table}%

 % Table generated by Excel2LaTeX from sheet 'dgp2-T10-alrv-d-power'
 \begin{table}[htbp]
 	\centering
 \tabcolsep=0.16cm
 \caption{EmpiricalPower of ${\cal E}_{n}(\mu_{0})$ - $\textrm{\bf DGP2}$}
 \vspace{0.4cm}
 \scalebox{0.88}{	\begin{tabular}{lcccccccc} \hline
 		& \multicolumn{4}{c}{$(N,T)=(100,250)$} & \multicolumn{4}{c}{$(N,T)=(500,500)$} \\ \hline
 		$\mu_{0}$ & 0.30  & 0.35  & 0.40  & 0.45  & 0.30  & 0.35  & 0.40  & 0.45 \\ \hline
 		& \multicolumn{4}{c}{h=1}       & \multicolumn{4}{c}{h=1} \\
 		$\beta_{2}$=0.10 & 0.204 & 0.222 & 0.269 & 0.361 & 0.283 & 0.348 & 0.469 & 0.677 \\
 		$\beta_{2}$=0.20 & 0.480 & 0.600 & 0.736 & 0.879 & 0.808 & 0.917 & 0.982 & 1.000 \\
 		$\beta_{2}$=0.30 & 0.824 & 0.916 & 0.976 & 0.996 & 0.994 & 0.999 & 1.000 & 1.000 \\
 		$\beta_{2}$=0.35 & 0.924 & 0.973 & 0.995 & 0.999 & 1.000 & 1.000 & 1.000 & 1.000 \\
 		$\beta_{2}$=0.40 & 0.977 & 0.996 & 1.000 & 1.000 & 1.000 & 1.000 & 1.000 & 1.000 \\
 		$\beta_{2}$=0.45 & 0.994 & 0.999 & 1.000 & 1.000 & 1.000 & 1.000 & 1.000 & 1.000 \\
 		$\beta_{2}$=0.50 & 0.999 & 1.000 & 1.000 & 1.000 & 1.000 & 1.000 & 1.000 & 1.000 \\
 		$\beta_{2}$=0.55 & 1.000 & 1.000 & 1.000 & 1.000 & 1.000 & 1.000 & 1.000 & 1.000 \\
 		$\beta_{2}$=0.60 & 1.000 & 1.000 & 1.000 & 1.000 & 1.000 & 1.000 & 1.000 & 1.000 \\
 		&       & h=4   &       &       &       & h=4   &       &  \\
 		$\beta_{2}$=0.10 & 0.187 & 0.189 & 0.216 & 0.242 & 0.233 & 0.271 & 0.325 & 0.417 \\
 		$\beta_{2}$=0.20 & 0.346 & 0.399 & 0.488 & 0.611 & 0.554 & 0.665 & 0.794 & 0.910 \\
 		$\beta_{2}$=0.30 & 0.592 & 0.689 & 0.798 & 0.898 & 0.892 & 0.955 & 0.989 & 0.998 \\
 		$\beta_{2}$=0.35 & 0.726 & 0.820 & 0.903 & 0.963 & 0.966 & 0.992 & 0.999 & 1.000 \\
 		$\beta_{2}$=0.40 & 0.827 & 0.901 & 0.960 & 0.987 & 0.990 & 0.999 & 1.000 & 1.000 \\
 		$\beta_{2}$=0.45 & 0.906 & 0.959 & 0.986 & 0.996 & 0.998 & 1.000 & 1.000 & 1.000 \\
 		$\beta_{2}$=0.50 & 0.950 & 0.982 & 0.995 & 0.999 & 1.000 & 1.000 & 1.000 & 1.000 \\
 		$\beta_{2}$=0.55 & 0.978 & 0.993 & 0.999 & 1.000 & 1.000 & 1.000 & 1.000 & 1.000 \\
 		$\beta_{2}$=0.60 & 0.991 & 0.998 & 1.000 & 1.000 & 1.000 & 1.000 & 1.000 & 1.000 \\
 		&       & h=12  &       &       &       & h=12  &       &  \\
 		$\beta_{2}$=0.10 & 0.184 & 0.190 & 0.211 & 0.243 & 0.226 & 0.257 & 0.310 & 0.404 \\
 		$\beta_{2}$=0.20 & 0.342 & 0.397 & 0.480 & 0.591 & 0.557 & 0.665 & 0.789 & 0.901 \\
 		$\beta_{2}$=0.30 & 0.579 & 0.675 & 0.780 & 0.879 & 0.881 & 0.948 & 0.983 & 0.997 \\
 		$\beta_{2}$=0.35 & 0.707 & 0.799 & 0.887 & 0.948 & 0.957 & 0.986 & 0.997 & 1.000 \\
 		$\beta_{2}$=0.40 & 0.804 & 0.885 & 0.946 & 0.982 & 0.988 & 0.998 & 1.000 & 1.000 \\
 		$\beta_{2}$=0.45 & 0.887 & 0.943 & 0.978 & 0.994 & 0.997 & 1.000 & 1.000 & 1.000 \\
 		$\beta_{2}$=0.50 & 0.934 & 0.973 & 0.992 & 0.998 & 0.999 & 1.000 & 1.000 & 1.000 \\
 		$\beta_{2}$=0.55 & 0.968 & 0.990 & 0.998 & 1.000 & 1.000 & 1.000 & 1.000 & 1.000 \\
 		$\beta_{2}$=0.60 & 0.984 & 0.996 & 0.999 & 1.000 & 1.000 & 1.000 & 1.000 & 1.000 \\
 		&       & h=24  &       &       &       & h=24  &       &  \\
 		$\beta_{2}$=0.10 & 0.180 & 0.193 & 0.203 & 0.229 & 0.230 & 0.259 & 0.311 & 0.401 \\
 		$\beta_{2}$=0.20 & 0.328 & 0.378 & 0.451 & 0.558 & 0.547 & 0.649 & 0.770 & 0.889 \\
 		$\beta_{2}$=0.30 & 0.551 & 0.647 & 0.756 & 0.856 & 0.871 & 0.944 & 0.983 & 0.998 \\
 		$\beta_{2}$=0.35 & 0.673 & 0.770 & 0.863 & 0.935 & 0.953 & 0.985 & 0.997 & 1.000 \\
 		$\beta_{2}$=0.40 & 0.771 & 0.858 & 0.935 & 0.975 & 0.984 & 0.997 & 1.000 & 1.000 \\
 		$\beta_{2}$=0.45 & 0.860 & 0.928 & 0.973 & 0.993 & 0.997 & 0.999 & 1.000 & 1.000 \\
 		$\beta_{2}$=0.50 & 0.922 & 0.966 & 0.988 & 0.997 & 1.000 & 1.000 & 1.000 & 1.000 \\
 		$\beta_{2}$=0.55 & 0.956 & 0.984 & 0.996 & 0.999 & 1.000 & 1.000 & 1.000 & 1.000 \\
 		$\beta_{2}$=0.60 & 0.979 & 0.996 & 0.999 & 1.000 & 1.000 & 1.000 & 1.000 & 1.000 \\
 	\end{tabular}}
 	\label{tab:addlabel}%
 \end{table}%

\section{Application: Can global inflation improve country specific inflation forecasts?}
We illustrate the use of our methodology through an application to inflation forecasts. More specifically, we are interested in assessing whether country specific inflation forecasts may benefit from pooling inflation data from multiple countries via a proxy for global inflation. This is an issue that has been recently investigated in \cite{cm2010} (CM2010) and our specific encompassing based test provides an ideal framework for formally {\it testing} such a phenomenon as opposed to a descriptive comparison of out of sample MSEs.
 
Here we restrict our attention to a global inflation factor constructed from observables rather than estimated through a factor model. For this purpose and following CM2010 we proxy global inflation via the cross-country average of individual country specific inflation series. In CM2010 the authors documented substantial accuracy gains for country specific inflation forecasts when a global inflation factor is included as a predictor within a conventional autoregressive specification. This regardless of whether one uses an estimated global inflation factor or a simple average of country specific inflation rates.  Given the well documented challenges for obtaining reliable inflation forecasts the findings in CM2010 spurred an important agenda around the globalisation of inflation and its drivers (see e.g., \cite{fk2019}, \cite{ms2012}, \cite{als2019}).

Here we follow closely the forecasting specifications considered in CM2010 with the aim of assessing whether country specific inflation forecasts based on an autoregressive model augmented with an additional global inflation proxy are encompassed within the forecasts implied by the purely autoregressive process so that our null hypothesis is that global factors do not contribute to improved country-specific inflation forecasts. 
It is also noteworthy to point out that the robustness of our methodology to the persistence properties of predictors is of particular relevance here due to the widely documented persistent nature of inflation time series. 

In what follows we let $\pi_{i,t}^{h} = (400/h)\ln(P_{i,t}/P_{i,t-h})$ denote country $i's$ h-quarter-annualized inflation 
in the price level $P_{i,t}$ and 
%$\pi_{t+h|t}^{h}$ its h-quarter ahead forecast
we consider the following two nested forecasting models for each country
 \begin{align}
 \pi_{i,t}^{h} & =  \beta_{01} + \sum_{j=0}^{p_{i,1}}\beta_{1j} \pi_{i,t-h-j}^{1} + w_{t,h}   \label{eq:eq22}  \\
 \pi_{i,t}^{h} & =  \beta_{02} + \sum_{j=0}^{p_{i,1}}\beta_{2j} \pi_{i,t-h-j}^{1} + \sum_{j=0}^{p_{2}}\gamma_{j} \pi_{t-h-j}^{g} + u_{t,h}  
   \label{eq:eq23} 
 \end{align}
where $\pi_{t}^{g}$ refers to our measure of global inflation. Note that the predictors in the right hand side of (\ref{eq:eq22})-(\ref{eq:eq23}) are quarter-on-quarter inflation rates e.g., for $j=0$ $\pi_{i,t-h}^{1}=400*\ln(P_{i,t-h}/P_{i,t-h-1})$, for $j=1$, $\pi_{i,t-h-1}^{1}=400*\ln(P_{i,t-h-1}/P_{i,t-h-2})$ and so on so that our setting conforms with a {\it direct} approach of building h-steps ahead forecasts. 
 
Our empirical implementation is based on the {\it global inflation database} recently made available by the World Bank (see \cite{hko2021}) and covering the period 1970-2023. We focus our attention on  quarterly series of the Headline Consumer Price Index (data item $\textrm{HCPI_Q}$) and implement our tests using one year ahead forecasts of inflation (i.e., we set $h=4$). The recursive estimation of the competing models in (\ref{eq:eq22})-(\ref{eq:eq23}) is implemented using 25\% of the sample to initiate recursions. The lag lengths $p_{i,1}$ (country-specific) are selected using the BIC criterion while we fix $p_{2}$ to $p_{2}=4$ throughout. Note that this preliminary lag selection exercise for the autoregressive structure of order $p_{i,1}$ is implemented by fitting model (\ref{eq:eq22}) and choosing $p_{i,1}$ as the minimiser of the corresponding BIC criterion. Regarding our metric for global inflation we follow CM2010 and consider 
the average of the $\pi_{i,t-h-j}'$ across all countries included in the analysis (see Table 6). 

%As it is  customary in most databases the CPI data has not been seasonally adjusted. However, the year on year inflation rates $\pi_{t}^{h}$ as defined above are not expected to have any seasonal components due to the seasonal differencing transformation. Naturally, this may not be the case for the quarter-on-quarter changes characterising the predictors (e.g., $\pi_{i,t-h}=400\ln(P_{i,t-h}/P_{i,t-h-1})$) although such effects will typically be particularly mild to non-visible as can be observed from the time-series plots in Figure 1. Nevertheless, for robustness considerations we have also seasonally adjusted all series using the US-Census provided 
%X-13ARIMA-SEATS program and re-implemented all our tests on these adjusted series. As there were no qualitative differences in outcomes we have relegated these additional results to a supplementary appendix accompanying this paper.  

In a first instance we focus on comparing the forecasting performance of (\ref{eq:eq23}) versus (\ref{eq:eq22}) 
{\it descriptively} via their associated RMSEs. For this purpose the first column of Table 6 presents the ratios of the root mean-squared errors of the larger global inflation model to the smaller purely autoregressive model. A ratio that is less than 1 indicates preference for the global inflation model. 

% Table generated by Excel2LaTeX from sheet 'Fetched_all_Q'
\begin{table}[htbp]
	\centering
	\caption{Global versus Local inflation}
	\vspace{0.28cm}
	\begin{tabular}{lccc}
		& RMSE ratios & \multicolumn{2}{c}{${\cal E}_{n}(\mu_{0})$} \\ \hline
		& $\textrm{RMSE}^{g}/\textrm{RMSE}^{ar}$ & \multicolumn{1}{l}{$\mu_{0}=0.40$} & \multicolumn{1}{l}{$\mu_{0}=0.45$} \\ \hline
		usa   & 1.104 & 0.690 & 0.967 \\
		gbr   & \textbf{0.844} & \textbf{0.000} & \textbf{0.000} \\
		jpn   & 1.319 & 1.000 & 1.000 \\
		fra   & 1.016 & 0.939 & 0.649 \\
		deu   & 1.033 & 0.648 & 0.422 \\
		esp   & \textbf{0.891} & \textbf{0.000} & \textbf{0.000} \\
		ita   & \textbf{0.796} & \textbf{0.000} & \textbf{0.000} \\
		nld   & \textbf{0.996} & \textbf{0.008} & \textbf{0.000} \\
		lux   & \textbf{0.945} & \textbf{0.000} & \textbf{0.000} \\
		can   & \textbf{0.966} & \textbf{0.000} & \textbf{0.000} \\
		irl   & \textbf{0.932} & \textbf{0.000} & \textbf{0.000} \\
		fin   & 1.001 & 0.100 & \textbf{0.000} \\
		nzl   & \textbf{0.898} & \textbf{0.000} & \textbf{0.000} \\
		grc   & \textbf{0.958} & 0.111 & \textbf{0.000} \\
		prt   & \textbf{0.803} & \textbf{0.000} & \textbf{0.000} \\
		nor   & \textbf{0.944} & \textbf{0.000} & \textbf{0.000} \\
		kor   & \textbf{0.899} & \textbf{0.000} & \textbf{0.000} \\
		dnk   & 1.008 & 0.431 & \textbf{0.000} \\
		swe   & \textbf{0.884} & \textbf{0.000} & \textbf{0.000} \\
		aus   & \textbf{0.956} & \textbf{0.000} & \textbf{0.000} \\
		aut   & 1.046 & 0.113 & 0.032 \\
		bel   & \textbf{0.945} & \textbf{0.000} & \textbf{0.000} \\
		che   & \textbf{0.966} & \textbf{0.000} & \textbf{0.000} \\
	\end{tabular}%
	\label{tab:addlabel}%
\end{table}%

Although the ratios of RMSEs are less than 1 for most countries, indicating that global inflation plays an important role in improving country-specific forecasts of inflation this is not the case for the world's largest economies (USA (usa), Japan (jpn), France (fra) and Germany (deu)) with the exception of the United-Kingdom (gbr). 
With a ratio of 1.319 Japan comes across as the most extreme example where global inflation plays no role whatsoever in one-year ahead inflation forecast accuracy as measured by RMSEs. For the United-Kingdom on the other hand we note a ratio of 0.844 indicating a very substantial role played by global inflation. Our proposed test now allows us to formalise such descriptive results via a formal encompassing based hypothesis test. The last two columns of Table 6 present p-values associated with our ${\cal E}_{n}(\mu_{0})$ statistic across two magnitudes of the tuning parameter $\mu_{0}$. We first note an overall agreement between our test based outcomes and the 
previous RMSE cmparisons. We continue to observe for instance a p-value near 1 for Japan and a p-value near 0 for the United-Kingdom. The latter implies for instance that the pure autoregression based forecasts of UK inflation do not encompass the forecasts obtained through the global inflation augmented model. Differently put, global inflation matters significantly and UK inflation forecasts can be substantially improved by taking global inflation into consideration. In the case of Japan on the other hand, the forecasts from the augmented model are 
clearly encompasses within the pure autoregression based forecasts so that global inflation appears to play no role in enhancing inflation forecasts. 

%Overall these empirical findings highlight the potentially important role played by {\it network effects} when it %comes to modelling and forecasting inflation. More elaborate models that take advantage of such effects are %worthy of further investigation. The case of the United-Kingdom as opposed to the remaining largest economies is %also 
%noteworthy and may be worth exploring further. A similar outcome was also observed in Ciccarelli and Mojon (2010) %whose study covered the 1960-2008 period as opposed to our 1970-2023 coverage. Using the AR benchmark, the global %inflation model was found to be particularly powerful for the UK while much less so for Japan and the %United-States. 

\section{Concluding Remarks}

This paper has introduced a formal test of the forecast encompassing hypothesis tailored to have a well defined limiting distribution under nested model comparisons/combinations. It can be used for both single-period and longer-horizon forecasts and its asymptotic properties also remain valid when persistent predictors are included in the fitted models. The ability of our test to bypass variance degeneracy that plagues this type of nested model based inferences is particularly noteworthy. This is achieved through a simple device that requires introducing a tuning parameter in the formulation of our proposed test statistic which we think can be of independent interest 
for testing population moment restrictions. Although this tuning parameter may be perceived as an ad-hoc input it is not more so than say choosing a bandwidth parameter. Our formal local power analysis provides a very clear and tight range for selecting a suitable magnitude for it and whose usefulness is backed by a comprehensive simulation study. 

Our empirical findings have highlighted the potentially important role played by {\it network effects} when it comes to modelling and forecasting inflation. More elaborate models that take advantage of such effects are worthy of further investigation. The case of the United-Kingdom as opposed to the other largest economies is also 
noteworthy and may be worth exploring further. A similar outcome was observed in CM2010 whose study covered the 1960-2008 period as opposed to our 1970-2023 coverage. Using an autoregressive benchmark, the global inflation model was found to be particularly powerful for the UK while much less so for Japan and the United-States.

\newpage

\section*{Appendix}

\noindent
PROOF OF PROPOSITION 1. We write $\sqrt{n} \ \overline{d}_{n}$ in (\ref{eq:eq7}) as
\begin{align}
	\sqrt{n} \ \overline{d}_{n} & = A_{1n}(m_{0})+A_{2n}(m_{0})  \label{eq:eq24} 
\end{align}
where 
\begin{align}
A_{1n}(m_{0})  & = \dfrac{1}{\sqrt{n}} \sum (u_{t+h}^{2}-\sigma^{2}_{u})-\dfrac{1}{2}
\left(\dfrac{n}{m_{0}} \dfrac{1}{\sqrt{n}} \sum (u_{t+h}^{2}-\sigma^{2}_{u})+\dfrac{n}{n-m_{0}} \dfrac{1}{\sqrt{n}} \sum (u_{t+h}^{2}-\sigma^{2}_{u})\right), \label{eq:eq25}   \\
A_{2n}(m_{0}) & = \dfrac{1}{\sqrt{n}} \sum (\hat{e}_{1,t+h|t}^{2}-u_{t+h}^{2})  \nonumber \\
& -\dfrac{1}{2}
\left(\dfrac{n}{m_{0}} \dfrac{1}{\sqrt{n}} \sum (\hat{e}_{1,t+h|t}\hat{e}_{2,t+h|t}-u_{t+h}^{2})+\dfrac{n}{n-m_{0}} \dfrac{1}{\sqrt{n}} \sum (\hat{e}_{1,t+h|t}\hat{e}_{2,t+h|t}-u_{t+h}^{2})\right)
 \label{eq:eq26} 
\end{align}
with $\sigma^{2}_{u}=E[u_{t+h}^{2}]$ and all summations in (\ref{eq:eq25})-(\ref{eq:eq26}) understood to cover the same ranges as in (\ref{eq:eq6}). It now follows directly from {\bf A2} and repeated applications of the triangle inequality onto (\ref{eq:eq26}) that 
\begin{align}
	\sqrt{n} \ \overline{d}_{n} & = A_{1n}(m_{0})+o_{p}(1). 
	 \label{eq:eq27} 
\end{align}
Appealing to the FCLT in {\bf A1} and the continuous mapping theorem leads to
\begin{align}
\sqrt{n} \ \overline{d}_{n} \stackrel{\cal D}\rightarrow \dfrac{-(1-2\mu_{0})}{2\mu_{0}(1-\mu_{0})} \phi [W(\mu_{0})-\mu_{0}W(1)]  \label{eq:eq28} 
\end{align}
which for a fixed and given $\mu_{0}$ is equivalent to
\begin{align}
	\sqrt{n} \ \overline{d}_{n} \stackrel{\cal D}\rightarrow {\cal N}\left(0,\dfrac{(1-2\mu_{0})^{2}}{4 \mu_{0}(1-\mu_{0})} \  \phi^{2}\right).  \label{eq:eq29} 
\end{align}
Next, $\sqrt{n} \ \overline{d}_{n}/\hat{\omega}_{n} = (\sqrt{n} \ \overline{d}_{n}/\omega_{0})(\omega_{0}/\hat{\omega}_{n})$ with $\omega_{0}^{2}=(1-2\mu_{0})^{2}\phi^{2}/4\mu_{0}(1-\mu_{0})$ so that the result in (\ref{eq:eq14}) of Proposition 1 follows provided that $\omega_{0}/\hat{\omega}_{n}\stackrel{p}\rightarrow 1$ or equivalently $\hat{\omega}_{n}^{2}\stackrel{p}\rightarrow \omega_{0}^{2}$ which is in turn ensured by {\bf A1(ii)}, noting that
\begin{align}
\hat{\omega}_{n}^{2} & = \dfrac{(1-2\mu_{0})^{2}}{4\mu_{0}(1-\mu_{0})} \ \hat{\phi}_{n}^{2}+o_{p}(1).
 \label{eq:eq30} 
\end{align}
\hfill $\blacksquare$ \\

\noindent
PROOF OF PROPOSITION 2. From (\ref{eq:eq2}) and letting $\bm \theta_{10}=(\alpha_{02},\bm \delta^{\top})^{\top}$ we write
\begin{align}
\hat{e}_{1,t+h|t} & = u_{t+h}+{\bm \gamma}^{\top}{\bm x}_{2,t}-(\hat{\bm \theta}_{1,t}-\bm \theta_{10})^{\top} \widetilde{\bm x}_{1,t}  \label{eq:eq31} \\
\hat{e}_{2,t+h|t} & =  u_{t+h}-(\hat{\bm \theta}_{2,t}-\bm \theta_{20})^{\top} \tilde{\bm x}_{12,t}.
 \label{eq:eq32} 
\end{align}
\noindent
and consider the limiting behaviour of $A_{2n}(m_{0})$ as defined in (\ref{eq:eq26}). Using (\ref{eq:eq31}) with ${\bm \gamma}={\bm c}/T^{1/4}$ and recalling that $n \equiv (T-k_{0}-h+1)$ we have 
\begin{align}
\dfrac{1}{\sqrt{n}}\sum_{t=k_{0}}^{T-h}(\hat{e}_{1,t+h|t}^{2}-u_{t+h}^{2}) & = \dfrac{\sqrt{n}}{\sqrt{T}} {\bm c}^{\top} \
\dfrac{\sum {\bm x_{2,t} \bm x_{2,t}^{\top}}}{n} \ {\bm c} 
+\dfrac{\sqrt{n}}{\sqrt{T}} \sum T^{1/4}(\hat{\bm \theta}_{1,t}-\bm \theta_{0})^{\top} \dfrac{\widetilde{\bm x}_{1,t}  \widetilde{\bm x}_{1,t}^{\top}}{n}
T^{1/4}(\hat{\bm \theta}_{1,t}-\bm \theta_{0}) \nonumber \\
& - 2 \ \dfrac{\sqrt{n}}{\sqrt{T}} \sum T^{1/4}(\hat{\bm \theta}_{1,t}-\bm \theta_{0})^{\top}\dfrac{{\widetilde{\bm x}_{1,t} \bm x_{2,t}^{\top}}}{n} \ {\bm c} \nonumber \\
& - 2 \ \frac{1}{T^{1/4}} \sum T^{1/4}(\hat{\bm \theta}_{1,t}-\bm \theta_{0})^{\top}\dfrac{{\widetilde{\bm x}_{1,t} u_{t+h}}}{\sqrt{n}}+\frac{1}{T^{1/4}} 2 \ {\bm c}^{\top} \ \dfrac{\sum \bm x_{2,t}u_{t+h}}{\sqrt{n}},
\label{eq:eq33} 
\end{align}

\begin{align}
\dfrac{1}{\sqrt{n}}\sum_{t=k_{0}}^{m_{0}+k_{0}-1}(\hat{e}_{1,t+h|t}\hat{e}_{2,t+h|t}-u_{t+h}^{2}) & = 
\dfrac{1}{T^{1/4}}\dfrac{\sqrt{n}}{\sqrt{T}} \sum T^{1/4}(\hat{\bm \theta}_{1t}-\bm \theta_{10})^{\top} 
\dfrac{\widetilde{\bm x}_{1,t} \widetilde{\bm x}_{12,t}}{n} 
T^{1/2} (\hat{\bm \theta}_{2t}-\bm \theta_{20}) \nonumber \\
& - \dfrac{1}{T^{1/4}} \dfrac{\sqrt{n}}{\sqrt{T}}\sum T^{1/2} (\hat{\bm \theta}_{2t}-\bm \theta_{20})^{\top} 
\dfrac{\widetilde{\bm x}_{12,t} \bm x_{2,t}^{\top}}{n} \ {\bm c} \nonumber \\
& + \dfrac{1}{T^{1/4}} {\bm c}^{\top} \dfrac{\sum {\bm x}_{2,t}u_{t+h}}{\sqrt{n}}
- \sum (\hat{\bm \theta}_{2t}-\bm \theta_{20})^{\top} \dfrac{\widetilde{\bm x}_{12,t}u_{t+h}}{\sqrt{n}} \nonumber \\
& - \dfrac{1}{T^{1/4}} \sum T^{1/4} (\hat{\bm \theta}_{1t}-\bm \theta_{10})^{\top} \dfrac{\widetilde{\bm x}_{1,t}u_{t+h}}{\sqrt{n}}
\label{eq:eq34} 
\end{align}
\noindent 
and
\begin{align}
	\dfrac{1}{\sqrt{n}}\sum_{t=k_{0}+m_{0}}^{T-h}(\hat{e}_{1,t+h|t}\hat{e}_{2,t+h|t}-u_{t+h}^{2}) & = 
	\dfrac{1}{T^{1/4}}\dfrac{\sqrt{n}}{\sqrt{T}} \sum T^{1/4}(\hat{\bm \theta}_{1t}-\bm \theta_{10})^{\top} 
	\dfrac{\widetilde{\bm x}_{1,t} \widetilde{\bm x}_{12,t}}{n} 
	T^{1/2} (\hat{\bm \theta}_{2t}-\bm \theta_{20}) \nonumber \\
	& - \dfrac{1}{T^{1/4}} \dfrac{\sqrt{n}}{\sqrt{T}}\sum T^{1/2} (\hat{\bm \theta}_{2t}-\bm \theta_{20})^{\top} 
	\dfrac{\widetilde{\bm x}_{12,t} \bm x_{2,t}^{\top}}{n} \ {\bm c} \nonumber \\
	& + \dfrac{1}{T^{1/4}} {\bm c}^{\top} \dfrac{\sum {\bm x}_{2,t}u_{t+h}}{\sqrt{n}}
	- \sum (\hat{\bm \theta}_{2t}-\bm \theta_{20})^{\top} \dfrac{\widetilde{\bm x}_{12,t}u_{t+h}}{\sqrt{n}} \nonumber \\
	& - \dfrac{1}{T^{1/4}} \sum T^{1/4} (\hat{\bm \theta}_{1t}-\bm \theta_{10})^{\top} \dfrac{\widetilde{\bm x}_{1,t}u_{t+h}}{\sqrt{n}}
\label{eq:eq35} 
\end{align}
\noindent 
Next, we note that 
\begin{align}
T^{1/4} (\hat{\bm \theta}_{1,t}-\bm \theta_{0}) & =  \left(\dfrac{\sum_{s=1}^{t} \widetilde{\bm x}_{1,s-h}\widetilde{\bm x}_{1,s-h}^{\top}}{n}\right)^{-1} \dfrac{\sum_{s=1}^{t}  \widetilde{\bm x}_{1,s-h} \ {\bm x}_{2,s}}{n} \ {\bm c} \nonumber \\
& + \dfrac{1}{n^{1/4}} \left( \dfrac{T^{1/4}}{n^{1/4}} \ \left(\dfrac{\sum_{s=1}^{t} \widetilde{\bm x}_{1,s-h}\widetilde{\bm x}_{1,s-h}^{\top}}{n}\right)^{-1} \dfrac{\sum_{s=1}^{t}  \widetilde{\bm x}_{1,s-h} u_{sh}}{\sqrt{n}}\right)
\label{eq:eq36} 
\end{align}
so that assumptions {\bf B1-B2} ensure that 
\begin{align}
\sup_{r}\left|
T^{1/4}(\hat{\bm \theta}_{1,[Tr]}-\bm \theta_{10})-\ {\bm Q}_{11}^{-1}{\bm Q}_{12} \ \bm c
\right| & = o_{p}(1) \label{eq:eq37} 
\end{align}
and similarly
\begin{align}
\sup_{r}\left|
T^{1/2}(\hat{\bm \theta}_{2,[Tr]}-\bm \theta_{20})\right| & = O_{p}(1). \label{eq:eq38} 
\end{align}
\noindent
Using (\ref{eq:eq37})-(\ref{eq:eq38}) in (\ref{eq:eq33})-(\ref{eq:eq35}) we can next write
\begin{align}
	\dfrac{1}{\sqrt{n}}\sum_{t=k_{0}}^{T-h}(\hat{e}_{1,t+h|t}^{2}-u_{t+h}^{2}) & = \dfrac{\sqrt{n}}{\sqrt{T}} {\bm c}^{\top} \
	\dfrac{\sum {\bm x_{2,t} \bm x_{2,t}^{\top}}}{n} \ {\bm c} 
	+\dfrac{\sqrt{n}}{\sqrt{T}} \sum T^{1/4}(\hat{\bm \theta}_{1,t}-\bm \theta_{0})^{\top} \dfrac{\widetilde{\bm x}_{1,t}  \widetilde{\bm x}_{1,t}^{\top}}{n}
	T^{1/4}(\hat{\bm \theta}_{1,t}-\bm \theta_{0}) \nonumber \\
	& - 2 \ \dfrac{\sqrt{n}}{\sqrt{T}} \sum T^{1/4}(\hat{\bm \theta}_{1,t}-\bm \theta_{0})^{\top}\dfrac{{\widetilde{\bm x}_{1,t} \bm x_{2,t}^{\top}}}{n} \ {\bm c} + O_{p}(T^{-1/4}),
	\label{eq:eq39} 
\end{align}
while 
\begin{align}
\dfrac{1}{\sqrt{n}}\sum_{t=k_{0}}^{m_{0}+k_{0}-1}(\hat{e}_{1,t+h|t}\hat{e}_{2,t+h|t}-u_{t+h}^{2}) & =
O_{p}(T^{-1/4})	
	\label{eq:eq40}
\end{align}
\noindent and 
\begin{align}
	\dfrac{1}{\sqrt{n}}\sum_{t=k_{0}+m_{0}}^{T-h}(\hat{e}_{1,t+h|t}\hat{e}_{2,t+h|t}-u_{t+h}^{2}) & =
	O_{p}(T^{-1/4}).
		\label{eq:eq41}	
\end{align}
so that under this local alternative scenario the limiting behaviour of $A_{2n}(m_{0})$ is 
driven by its first component in the right hand side of (\ref{eq:eq26}). Finally, combining the results in (\ref{eq:eq37})-(\ref{eq:eq38}) within (\ref{eq:eq39}) and appealing also to assumptions {\bf B1-B2} 
gives $\sqrt{n} \ \overline{d}_{n}\stackrel{\cal D}\rightarrow {\cal N}(0,\omega^{2}_{0})+\sqrt{1-\pi_{0}} \ {\bm c}'(\bm Q_{22}-\bm Q_{21} \bm Q_{11}^{-1} \bm Q_{12}){\bm c}$ which upon normalization with $\hat{\omega}_{n}$ leads to the stated result in (\ref{eq:eq15}). \hfill $\blacksquare$ \\

\noindent
PROOF OF PROPOSITION 3. We proceed with the following alternative normalizations applied to the components of $A_{2n}(m_{0})$ as defined in (\ref{eq:eq26})
\begin{align}
	\dfrac{1}{\sqrt{n}}\sum_{t=k_{0}}^{T-h}(\hat{e}_{1,t+h|t}^{2}-u_{t+h}^{2}) & = \left(\dfrac{n}{T}\right)^{\frac{1}{2}+\alpha} {\bm c}^{\top} \
	\dfrac{\sum {\bm x_{2,t} \bm x_{2,t}^{\top}}}{n^{1+\alpha}} \ {\bm c} \nonumber \\
	& +\left(\dfrac{n}{T}\right)^{\frac{1}{2}+\alpha} \sum T^{\frac{1}{4}+\frac{\alpha}{2}}(\hat{\bm \theta}_{1,t}-\bm \theta_{0})^{\top} \dfrac{\widetilde{\bm x}_{1,t}  \widetilde{\bm x}_{1,t}^{\top}}{n^{1+\alpha}}
	T^{\frac{1}{4}+\frac{\alpha}{2}} (\hat{\bm \theta}_{1,t}-\bm \theta_{0}) \nonumber \\
	& - 2 \ \left(\dfrac{n}{T}\right)^{\frac{1}{2}+\alpha} \sum T^{\frac{1}{4}+\frac{\alpha}{2}} (\hat{\bm \theta}_{1,t}-\bm \theta_{0})^{\top}\dfrac{{\widetilde{\bm x}_{1,t} \bm x_{2,t}^{\top}}}{n^{1+\alpha}} \ {\bm c} \nonumber \\
	& - 2 \  \left(\dfrac{n}{T}\right)^{\frac{\alpha}{2}} \frac{1}{T^{1/4}} \sum T^{\frac{1}{4}+\frac{\alpha}{2}} (\hat{\bm \theta}_{1,t}-\bm \theta_{0})^{\top}\dfrac{{\widetilde{\bm x}_{1,t} u_{t+h}}}{n^{\frac{1+\alpha}{2}}} \nonumber \\
	& +\frac{1}{T^{1/4}} \left(\dfrac{n}{T}\right)^{\frac{\alpha}{2}} 2 \ {\bm c}^{\top} \ \dfrac{\sum \bm x_{2,t}u_{t+h}}{n^{\frac{1+\alpha}{2}}},
	\label{eq:eq42} 
\end{align}

\begin{align}
	\dfrac{1}{\sqrt{n}}\sum_{t=k_{0}}^{m_{0}+k_{0}-1}(\hat{e}_{1,t+h|t}\hat{e}_{2,t+h|t}-u_{t+h}^{2}) & = 
	\dfrac{1}{T^{1/4}}\left(\dfrac{n}{T}\right)^{\frac{1}{2}+\alpha} \sum T^{\frac{1}{4}+\frac{\alpha}{2}} (\hat{\bm \theta}_{1t}-\bm \theta_{10})^{\top} 
	\dfrac{\widetilde{\bm x}_{1,t} \widetilde{\bm x}_{12,t}}{n^{1+\alpha}} 
	T^{\frac{1+\alpha}{2}} (\hat{\bm \theta}_{2t}-\bm \theta_{20}) \nonumber \\
	& - \dfrac{1}{T^{1/4}} \left(\dfrac{n}{T}\right)^{\frac{1}{2}+\alpha} \sum T^{\frac{1+\alpha}{2}} (\hat{\bm \theta}_{2t}-\bm \theta_{20})^{\top} 
	\dfrac{\widetilde{\bm x}_{12,t} \bm x_{2,t}^{\top}}{n^{1+\alpha}} \ {\bm c} \nonumber \\
	& + \dfrac{1}{T^{1/4}} \left(\dfrac{n}{T}\right)^{\frac{\alpha}{2}} 	
	{\bm c}^{\top} \dfrac{\sum {\bm x}_{2,t}u_{t+h}}{n^{\frac{1+\alpha}{2}}} \nonumber \\
	& - \dfrac{1}{\sqrt{T}} \left(\dfrac{n}{T}\right)^{\frac{\alpha}{2}} \sum T^{\frac{1+\alpha}{2}}  (\hat{\bm \theta}_{2t}-\bm \theta_{20})^{\top} \dfrac{\widetilde{\bm x}_{12,t}u_{t+h}}{n^{\frac{1+\alpha}{2}}} \nonumber \\
	& - \dfrac{1}{T^{1/4}} \left(\dfrac{n}{T}\right)^{\frac{\alpha}{2}}  \sum T^{\frac{1}{4}+\frac{\alpha}{2}} (\hat{\bm \theta}_{1t}-\bm \theta_{10})^{\top} \dfrac{\widetilde{\bm x}_{1,t}u_{t+h}}{n^{\frac{1+\alpha}{2}}}
	\label{eq:eq43} 
\end{align}
\noindent
and
\begin{align}
	\dfrac{1}{\sqrt{n}}\sum_{t=k_{0}+m_{0}}^{T-h}(\hat{e}_{1,t+h|t}\hat{e}_{2,t+h|t}-u_{t+h}^{2}) & = 
	\dfrac{1}{T^{1/4}}\left(\dfrac{n}{T}\right)^{\frac{1}{2}+\alpha} \sum T^{\frac{1}{4}+\frac{\alpha}{2}} (\hat{\bm \theta}_{1t}-\bm \theta_{10})^{\top} 
	\dfrac{\widetilde{\bm x}_{1,t} \widetilde{\bm x}_{12,t}}{n^{1+\alpha}} 
	T^{\frac{1+\alpha}{2}} (\hat{\bm \theta}_{2t}-\bm \theta_{20}) \nonumber \\
	& - \dfrac{1}{T^{1/4}} \left(\dfrac{n}{T}\right)^{\frac{1}{2}+\alpha} \sum T^{\frac{1+\alpha}{2}} (\hat{\bm \theta}_{2t}-\bm \theta_{20})^{\top} 
	\dfrac{\widetilde{\bm x}_{12,t} \bm x_{2,t}^{\top}}{n^{1+\alpha}} \ {\bm c} \nonumber \\
	& + \dfrac{1}{T^{1/4}} \left(\dfrac{n}{T}\right)^{\frac{\alpha}{2}} 	
	{\bm c}^{\top} \dfrac{\sum {\bm x}_{2,t}u_{t+h}}{n^{\frac{1+\alpha}{2}}} \nonumber \\
	& - \dfrac{1}{\sqrt{T}} \left(\dfrac{n}{T}\right)^{\frac{\alpha}{2}} \sum T^{\frac{1+\alpha}{2}}  (\hat{\bm \theta}_{2t}-\bm \theta_{20})^{\top} \dfrac{\widetilde{\bm x}_{12,t}u_{t+h}}{n^{\frac{1+\alpha}{2}}} \nonumber \\
	& - \dfrac{1}{T^{1/4}} \left(\dfrac{n}{T}\right)^{\frac{\alpha}{2}}  \sum T^{\frac{1}{4}+\frac{\alpha}{2}} (\hat{\bm \theta}_{1t}-\bm \theta_{10})^{\top} \dfrac{\widetilde{\bm x}_{1,t}u_{t+h}}{n^{\frac{1+\alpha}{2}}}.
	\label{eq:eq44} 
\end{align}

Next, assumptions {\bf C1-C2} ensure that 
\begin{align}
	\sup_{r}\left| T^{\frac{1}{4}+\frac{\alpha}{2}}
	(\hat{\bm \theta}_{1,[Tr]}-\bm \theta_{10})-\ ({\bm V^{b}}_{11})^{-1}{\bm V^{b}}_{12} \ \bm c
	\right| & = o_{p}(1) \label{eq:eq45} 
\end{align}
and similarly, 
\begin{align}
	\sup_{r}\left|
T^{\frac{1+\alpha}{2}} (\hat{\bm \theta}_{2,[Tr]}-\bm \theta_{20})\right| & = O_{p}(1). \label{eq:eq46} 
\end{align}
so that (\ref{eq:eq42})-(\ref{eq:eq44}) satisfy

\begin{align}
	\dfrac{1}{\sqrt{n}}\sum_{t=k_{0}}^{T-h}(\hat{e}_{1,t+h|t}^{2}-u_{t+h}^{2}) & = \left(\dfrac{n}{T}\right)^{\frac{1}{2}+\alpha} {\bm c}^{\top} \
	\dfrac{\sum {\bm x_{2,t} \bm x_{2,t}^{\top}}}{n^{1+\alpha}} \ {\bm c} \nonumber \\
	& +\left(\dfrac{n}{T}\right)^{\frac{1}{2}+\alpha} \sum T^{\frac{1}{4}+\frac{\alpha}{2}}(\hat{\bm \theta}_{1,t}-\bm \theta_{0})^{\top} \dfrac{\widetilde{\bm x}_{1,t}  \widetilde{\bm x}_{1,t}^{\top}}{n^{1+\alpha}}
	T^{\frac{1}{4}+\frac{\alpha}{2}} (\hat{\bm \theta}_{1,t}-\bm \theta_{0}) \nonumber \\
	& - 2 \ \left(\dfrac{n}{T}\right)^{\frac{1}{2}+\alpha} \sum T^{\frac{1}{4}+\frac{\alpha}{2}} (\hat{\bm \theta}_{1,t}-\bm \theta_{0})^{\top}\dfrac{{\widetilde{\bm x}_{1,t} \bm x_{2,t}^{\top}}}{n^{1+\alpha}} \ {\bm c} + O_{p}(T^{-1/4}),
	\label{eq:eq47} 
\end{align}

\begin{align}
\dfrac{1}{\sqrt{n}}\sum_{t=k_{0}}^{m_{0}+k_{0}-1}(\hat{e}_{1,t+h|t}\hat{e}_{2,t+h|t}-u_{t+h}^{2}) & =
	O_{p}(T^{-1/4})	
	\label{eq:eq48}
\end{align}
\noindent and 
\begin{align}
\dfrac{1}{\sqrt{n}}\sum_{t=k_{0}+m_{0}}^{T-h}(\hat{e}_{1,t+h|t}\hat{e}_{2,t+h|t}-u_{t+h}^{2}) & =
	O_{p}(T^{-1/4}).
	\label{eq:eq49}	
\end{align}
Since we have
\begin{align}
A_{2n}(m_{0}) & = \left(\dfrac{n}{T}\right)^{\frac{1}{2}+\alpha} {\bm c}^{\top} \
\dfrac{\sum {\bm x_{2,t} \bm x_{2,t}^{\top}}}{n^{1+\alpha}} \ {\bm c} \nonumber \\
& +\left(\dfrac{n}{T}\right)^{\frac{1}{2}+\alpha} \sum T^{\frac{1}{4}+\frac{\alpha}{2}}(\hat{\bm \theta}_{1,t}-\bm \theta_{0})^{\top} \dfrac{\widetilde{\bm x}_{1,t}  \widetilde{\bm x}_{1,t}^{\top}}{n^{1+\alpha}}
T^{\frac{1}{4}+\frac{\alpha}{2}} (\hat{\bm \theta}_{1,t}-\bm \theta_{0}) \nonumber \\
& - 2 \ \left(\dfrac{n}{T}\right)^{\frac{1}{2}+\alpha} \sum T^{\frac{1}{4}+\frac{\alpha}{2}} (\hat{\bm \theta}_{1,t}-\bm \theta_{0})^{\top}\dfrac{{\widetilde{\bm x}_{1,t} \bm x_{2,t}^{\top}}}{n^{1+\alpha}} \ {\bm c} + O_{p}(T^{-1/4}),
\label{eq:eq50} 
\end{align}
appealing to assumptions {\bf C1-C2} and (\ref{eq:eq45})-(\ref{eq:eq50}) used within (\ref{eq:eq50}) leads to the stated result. \hfill $\blacksquare$

\end{document}